\pdfoutput=1
\RequirePackage{ifpdf}
\ifpdf
\documentclass[pdftex]{sigma}
\else
\documentclass{sigma}
\fi

\numberwithin{equation}{section}

\newtheorem{Theorem}{Theorem}[section]
\newtheorem{Lemma}[Theorem]{Lemma}
 { \theoremstyle{definition}
\newtheorem{Definition}[Theorem]{Definition}
\newtheorem{Remark}[Theorem]{Remark} }

\newcommand{\sth}{\scriptscriptstyle \rm I\hspace{-1pt}I\hspace{-1pt}I}
\newcommand{\so}{\scriptscriptstyle \rm I}
\newcommand{\st}{\scriptscriptstyle \rm I\hspace{-1pt}I}
\newcommand{\bu}{\bar u}
\newcommand{\bv}{\bar v}
\newcommand{\bet}{\bar\eta}
\newcommand{\bw}{\bar w}
\newcommand{\bxi}{\bar\xi}
\newcommand{\BV}{\mathbb{B}}
\newcommand{\str}{\mathop{\rm str}}
\newcommand{\bT}{\mathbb{T}}

\begin{document}


\renewcommand{\thefootnote}{$\star$}

\newcommand{\arXivNumber}{1605.06419}

\renewcommand{\PaperNumber}{099}

\FirstPageHeading

\ShortArticleName{Multiple Actions of the Monodromy Matrix in $\mathfrak{gl}(2|1)$-Invariant Integrable Models}

\ArticleName{Multiple Actions of the Monodromy Matrix\\
 in $\boldsymbol{\mathfrak{gl}(2|1)}$-Invariant Integrable Models\footnote{This paper is a~contribution to the Special Issue on Recent Advances in Quantum Integrable Systems. The full collection is available at \href{http://www.emis.de/journals/SIGMA/RAQIS2016.html}{http://www.emis.de/journals/SIGMA/RAQIS2016.html}}}

\Author{Arthur HUTSALYUK~$^{\dag^1}$, Andrii LIASHYK~$^{\dag^2\dag^3}$, Stanislav Z.~PAKULIAK~$^{\dag^4\dag^1}$,\newline
Eric RAGOUCY~$^{\dag^5}$ and Nikita A.~SLAVNOV~$^{\dag^6}$}

\AuthorNameForHeading{A.~Hutsalyuk, A.~Liashyk, S.Z.~Pakuliak, E.~Ragoucy and N.A.~Slavnov}

\Address{$^{\dag^1}$~Moscow Institute of Physics and Technology, Dolgoprudny, Moscow region, Russia}
\EmailDD{\href{mailto:hutsalyuk@gmail.com}{hutsalyuk@gmail.com}}

\Address{$^{\dag^2}$~Bogoliubov Institute for Theoretical Physics, NAS of Ukraine, Kyiv, Ukraine}
\EmailDD{\href{mailto:a.liashyk@gmail.com}{a.liashyk@gmail.com}}

\Address{$^{\dag^3}$~National Research University Higher School of Economics, Russia}

\Address{$^{\dag^4}$~Laboratory of Theoretical Physics, JINR, Dubna, Moscow region, Russia}
\EmailDD{\href{mailto:stanislav.pakuliak@jinr.ru}{stanislav.pakuliak@jinr.ru}}

\Address{$^{\dag^5}$~Laboratoire de Physique Th\'eorique LAPTh, CNRS and USMB, Annecy-le-Vieux, France}
\EmailDD{\href{mailto:eric.ragoucy@lapth.cnrs.fr}{eric.ragoucy@lapth.cnrs.fr}}

\Address{$^{\dag^6}$~Steklov Mathematical Institute of Russian Academy of Sciences, Moscow, Russia}
\EmailDD{\href{mailto:nslavnov@mi.ras.ru}{nslavnov@mi.ras.ru}}

\ArticleDates{Received June 24, 2016, in f\/inal form October 03, 2016; Published online October 08, 2016}

\Abstract{We study $\mathfrak{gl}(2|1)$ symmetric integrable models solvable by the nested algebraic Bethe ansatz. Using explicit formulas for the Bethe vectors we derive the actions of the monodromy matrix entries onto these vectors. We show that the result of these actions is a~f\/inite linear combination of Bethe vectors. The obtained formulas open a way for studying scalar products of Bethe vectors.}


\Classification{82B23; 81R12; 81R50; 17B80}

\renewcommand{\thefootnote}{\arabic{footnote}}
\setcounter{footnote}{0}

\section{Introduction}
Algebraic Bethe ansatz is one of the most famous applications of the quantum inverse scattering method. It was developed at the end of the 70s of the last century by the Leningrad school~\mbox{\cite{FadSklT79,FadTak84}}. From a mathematical point of view the method consists of the study of the highest weight representations of some Hopf algebra that depends on the model under considera\-tion~\mbox{\cite{Dri85,Jim85,KhoPak08}}. This method allows one to f\/ind eigenvectors of the transfer matrix (a generating function of the commuting Hamiltonians of the quantum models), leading to a diagonalization of the physical Hamiltonians. The obtained determinant representations for scalar products of the Bethe vectors~\cite{Kor82,Sla89} allow then to calculate the form factors in various integrable systems~\cite{KitKozlMaiSlavnTer,KitMaiT99,KorIzEss}. Using the form factor representations, many interesting physical results were obtained during the last few years~\cite{CauCS07,CauHM05, GohKS04,GohKS05,KitKMST09b,KitMST02,KitMT00,PerSCHMWA06,PerSCHMWA07,SeeBGK07}.

The results listed above mostly concerned the models with $\mathfrak{gl}(2)$ symmetry or its $q$-defor\-ma\-tion. An important problem remains open: the case of models based on the $\mathfrak{gl}(N)$ algebras ($N>2$). These models are quite more involved, therefore they are less studied. The models with higher rank symmetries were considered within the framework of a {\it nested Bethe ansatz} approach~\cite{Sath68,Sath75, Yang67}. The algebraic version of this method was developed in the works \cite{Kul81,KulRes83, KulRes82, KulS80}. Later on, new methods to construct Bethe vectors in the models of higher rank symmetries were proposed in \cite{KhoPak08, TarVar13}. A generalization to the case of superalgebras $\mathfrak{gl}(m|n)$ was given in \cite{BelR08}.

Recently, the models with $\mathfrak{gl}(3)$ symmetries were studied more deeply by the nested algebraic Bethe ansatz in the works
\cite{BelPakR10,BelPakRS12c,BelPRS12a,BelPakRS12d,PakRS15d,PakRS15c,BelPakRS14a,Whe13,Whe12}. There, determinant representations for scalar products of Bethe vectors and form factors of local operators were obtained. Some of these results were generalized to the models with $q$-deformed $\mathfrak{gl}(3)$ symmetry \cite{PRStrigoB,PRStrigoA,Sla15a}.

In this paper, we calculate multiple actions of the monodromy matrix entries onto Bethe vectors in the models with $\mathfrak{gl}(2|1)$ symmetry. This is a part of a larger program devoted to the study of quantum integrable models based on $\mathfrak{gl}(m|n)$ superalgebras. This program was initiated in~\cite{BelR08} and continued in~\cite{PakRS16a}, where the Bethe vectors for these models were constructed. The calculation of the monodromy matrix entries multiple actions onto the Bethe vectors is the next step of our investigation. Further steps that include the calculation of the scalar products and form factors of the monodromy matrix entries will be reported elsewhere. They will lead eventually to determinant representations for form factors of local operators in the models of physical interest. In particular, we expect to obtain compact explicit formulas for form factors of local operators in the supersymmetric $t$-$J$ model \cite{And90,EssK92,FoeK93,For89,Sch87, ZhaR88} and the $U$-model of strongly correlated electrons \cite{BedF95, BraGLZ95} since these models are based on the $\mathfrak{gl}(1|2)$ superalgebra. Knowing form factors of local operators, one can then study correlation functions via a~summation of their form factor series. We expect to come back on these physical models in further works. We also hope that our results will be of some interest in the context of super-Yang--Mills theories, when studied in the integrable systems framework. Indeed in these theories, the general approach relies on a spin chain based on the $\mathfrak{psl}(2,2|4)$ superalgebra, but there are closed subsectors based on the $\mathfrak{gl}(1|2)$ or $\mathfrak{gl}(2|1)$ superalgebras. We believe that the present results will be a f\/irst step to understand these subsectors, and then the entire theory.

The paper is organized as follows. Section~\ref{S-DefNot} is devoted to the def\/initions and the notation. In Section~\ref{Mult-action} we present the main results of the paper: multiple actions of the monodromy matrix entries onto Bethe vectors. In Section~\ref{S-OSBV} we consider the action of the transfer matrix onto on-shell Bethe vectors. Sections~\ref{ilessj}--\ref{Induction} contain proofs of the formulas presented in Section~\ref{Mult-action}. Finally, the appendix gathers some identities for rational functions that we use in our proofs.

\section{Def\/initions}\label{S-DefNot}

We consider a normalized rational $R$-matrix
 \begin{gather}\label{R-mat}
 R(x,y)=\mathbb{I}+g(x,y)P, \qquad g(x,y)=\frac{c}{x-y},
 \end{gather}
where $c$ is a constant and $P$ is a graded permutation operator \cite{KulS80}. The $R$-matrix of $\mathfrak{gl}(2|1)$-based models acts in the tensor product $\mathbb{C}^{2|1}\otimes \mathbb{C}^{2|1}$, where $\mathbb{C}^{2|1}$ is the $\mathbb{Z}_2$-graded vector space with the grading $[1]=[2]=0$, $[3]=1$.
The $R$-matrix \eqref{R-mat} satisf\/ies the graded Yang--Baxter equation
\begin{gather}\label{Y-B}
R_{12}(x,y)R_{13}(x,z)R_{23}(y,z)=R_{23}(y,z)R_{13}(x,z)R_{12}(x,y)
\end{gather}
 written in the tensor product of graded spaces $\mathbb{C}^{2|1}\otimes \mathbb{C}^{2|1}\otimes \mathbb{C}^{2|1}$. The subscripts in \eqref{Y-B} show in which copies of the $\mathbb{C}^{2|1}$ space the $R$-matrix acts non-trivially. The monodromy matrix $T(u)$ is also graded according to the rule $[T_{ij}(u)]=[i]+[j]$. It satisf\/ies an intertwining relation ($RTT$ relation):
\begin{gather}\label{RTT}
R(u,v)\bigl(T(u)\otimes \mathbb{I}\bigr) \bigl(\mathbb{I}\otimes T(v)\bigr)= \bigl(\mathbb{I}\otimes T(v)\bigr)
\bigl( T(u)\otimes \mathbb{I}\bigr)R(u,v).
\end{gather}
Equation \eqref{RTT} holds in the tensor product of graded spaces $\mathbb{C}^{2|1}\otimes \mathbb{C}^{2|1}\otimes\mathcal{H}$, where $\mathcal{H}$ is a~Hilbert space of a Hamiltonian. It implies commutation relations between the monodromy matrix entries:
\begin{gather}\label{commTT}
[T_{ij}(u),T_{kl}(v)\}=(-1)^{[i]([k]+[l])+[k][l]}g(u,v)\big[T_{kj}(v)T_{il}(u)-T_{kj}(u)T_{il}(v)\big],
\end{gather}
where we introduced a graded commutator
\begin{gather*}
[T_{ij}(u),T_{kl}(v)\}= T_{ij}(u)T_{kl}(v) -(-1)^{([i]+[j])([k]+[l])} T_{kl}(v) T_{ij}(u).
\end{gather*}
The graded transfer matrix is def\/ined as the supertrace of the monodromy matrix
\begin{gather*}
\mathcal{T}(u)=\str T(u)= \sum_{i=1}^{3} (-1)^{[i]} T_{ii}(u).
\end{gather*}
It def\/ines an integrable system, due to the relation $[\mathcal{T}(u),\mathcal{T}(v)]=0$ which is implied by the relation \eqref{RTT}.

In order to make our formulas more compact we use several auxiliary functions and conventions on the notation. In addition to the functions $g(x,y)$ we also introduce the functions
\begin{gather*}
f(x,y)=1+g(x,y)=\frac{x-y+c}{x-y},\qquad h(x,y)=\frac{f(x,y)}{g(x,y)}=\frac{x-y+c}{c},\\
t(x,y)=\frac{g(x,y)}{h(x,y)}=\frac{c^2}{(x-y+c)(x-y)}.
\end{gather*}
The following obvious properties of these functions are useful
\begin{gather*}
 g(x,y)=-g(y,x),\qquad h(x,y)=\frac{1}{g(x,y-c)},\qquad f(x-c,y)=\frac{1}{f(y,x)}.
\end{gather*}

Below, we will permanently have to deal with sets of variables which will be denoted by a~bar: $\bar u$, $\bar v$, $\bet$ etc. Individual elements of the sets are denoted by subscripts and without a bar: $u_k$, $v_{\ell}$, $\eta_j$ etc. The notation $\bu\pm c$ means that all the elements of the set $\bu$ are shifted by~$\pm c$: $\bu\pm c=\{u_1\pm c,\dots,u_n\pm c\}$. As a rule, the number of elements in the sets is not shown explicitly; however we give these cardinalities in special comments to the formulas. Subsets of variables are denoted by roman or other subscripts, easily distinguishable from Latin indices used for individual element: $\bar u_{\so}$, $\bar v_{\st}$, $\bet_{\rm ii}$, $\bxi_{0}$ etc. For example, the notation $\bar u\Rightarrow\{\bar u_{\so},\bar u_{\st}\}$ means that the set $\bar u$ is divided into two disjoint subsets $\bar u_{\so}$ and $\bar u_{\st}$. We assume that the elements in every subset are ordered in such a way that the sequence of their subscripts is strictly increasing. For the union of two sets into another one we use the notation $\{ \bv, \bar z\}=\bar \xi$. Finally we use a special notation $\bar u_{j}$, $\bar v_{k}$ and so on for the sets $\bar u\setminus \{u_j\}$, $\bar v\setminus \{v_k\}$ etc.

In order to avoid excessively cumbersome formulas we use shorthand notation for products of functions depending on one or two variables. Namely, whenever such a function depends on a set of variables, this means that we deal with the product of this function with respect to the corresponding set, as follows
\begin{gather*}
 h(\bar u,v)=\prod_{u_j\in\bar u} h(u_j,v),\qquad
 g(x_k, \bxi_\ell)= \prod_{\substack{\xi_j\in\bxi\\ \xi_j\ne \xi_\ell}} g(x_k, \xi_j),\qquad
 f(\bar u_{\st},\bar u_{\so})=\prod_{u_j\in\bar u_{\st}}\prod_{u_k\in\bar u_{\so}} f(u_j,u_k).
\end{gather*}
This notation is also used for the product of commuting operators,
\begin{gather}\label{SH-prod-O}
T_{ij}(\bar u)=\prod_{u_k\in\bar u} T_{ij}(u_k), \qquad\text{if}\quad [i]+[j]=0,\quad \mod(2).
\end{gather}
One can easily see from the commutation relations \eqref{commTT} that in this case $[T_{ij}(u),T_{ij}(v)]=0$, and hence, the operator product \eqref{SH-prod-O} is well def\/ined. However, if $[i]+[j]=1$, then $[T_{ij}(u),T_{ij}(v)]\ne 0$, but we can introduce symmetric operator products
\begin{gather*} 
\bT_{j3}(\bv)= \frac{T_{j3}(v_1)\cdots T_{j3}(v_n)}{\prod\limits_{n\ge \ell>m\ge 1} h(v_\ell,v_m)},
\qquad \bT_{3j}(\bv)= \frac{T_{3j}(v_1)\cdots T_{3j}(v_n)}{\prod\limits_{n\ge \ell>m\ge 1} h(v_m,v_\ell)}, \qquad j=1,2.
\end{gather*}

In various formulas the Izergin determinant $K_n(\bar x|\bar y)$ appears\footnote{Note that by def\/inition this function depends on two sets of variables. Therefore, the convention on shorthand notations for the products is not applicable in this case.}. It is def\/ined for two sets $\bar x$ and $\bar y$ with common cardinality $\# \bar x=\# \bar y=n$,
\begin{gather}\label{K-def}
 K_n(\bar x|\bar y)=h(\bar x,\bar y) \prod_{\ell<m}^n g(x_\ell,x_m)g(y_m,y_\ell) \det_n [ t(x_i,y_j) ].
\end{gather}
We draw the readers attention that according to the convention on the shorthand nota\-tion~$h(\bar x,\bar y)$ in~\eqref{K-def} means the double product of the $h$-functions over all parameters $\bar x$ and $\bar y$. It is easy to see from def\/inition~\eqref{K-def} that $K_1(x|y)=g(x,y)$ and
\begin{gather}\label{K-K}
K_n(\bar x|\bar y+c)=(-1)^n\frac{K_n(\bar y|\bar x)}{f(\bar y,\bar x)}.
\end{gather}

\subsection{Bethe vectors}

For further calculation we need explicit formulas for $\mathfrak{gl}(2|1)$ Bethe vectors in terms of the monodromy matrix entries $T_{ij}(u)$. We recall that generically Bethe vectors are special polynomials in operators $T_{ij}(u)$ with $i\le j$ applied to the pseudovacuum vector~$\Omega$. This vector possesses the following properties:
\begin{gather}\label{vac}
T_{ii}(u)\Omega =\lambda_i(u)\Omega,\quad i=1,2,3, \qquad T_{ij}(u)\Omega =0,\quad 3\geq i>j\geq 1.
\end{gather}
Here $\lambda_i(u)$ are some scalar functions depending on a specif\/ic model. Below we will also use the ratios of these functions
\begin{gather*}
r_1(u)=\frac{\lambda_1(u)}{\lambda_2(u)}, \qquad r_3(u)=\frac{\lambda_3(u)}{\lambda_2(u)}.
\end{gather*}
We extend the convention on the shorthand notation to the products of the functions $\lambda_i$ and $r_k$. For example,
\begin{gather*}
\lambda_2(\bar z)=\prod_{z_j\in\bar z} \lambda_2(z_j),\qquad r_1(\bet_{\st})=\prod_{\eta_j\in\bet_{\st}}r_1(\eta_j).
\end{gather*}

We denote the Bethe vectors as $\BV_{a,b}(\bu;\bv)$. They depend on two sets of variables (Bethe parameters) $\bu=\{u_1,\dots,u_a\}$ and $\bv=\{v_1,\dots,v_b\}$, where $a,b=0,1,\dots$. Explicit representations for $\mathfrak{gl}(2|1)$ Bethe vectors were obtained in\footnote{The formulas for the Bethe vectors obtained in \cite{PakRS16a} dif\/fer from \eqref{Phi-expl1} and \eqref{Phi-expl2} by a normalization factor $\lambda_2(\bv)\lambda_2(\bu) f(\bv,\bu)$.}~\cite{PakRS16a}.

\begin{Definition}
For $\#\bu=a$ and $\#\bv=b$ def\/ine a Bethe vector
\begin{gather}\label{Phi-expl1}
\BV_{a,b}(\bu;\bv)=\sum g(\bv_{\so},\bu_{\so}) \frac{ f(\bu_{\so},\bu_{\st}) g(\bv_{\st},\bv_{\so})h(\bu_{\so},\bu_{\so})}
{\lambda_2(\bu)\lambda_2(\bv_{\st})f(\bv,\bu)} \bT_{13}(\bu_{\so}) T_{12}(\bu_{\st}) \bT_{23}(\bv_{\st})\Omega.
\end{gather}
Here the sum is taken over partitions $\bv\Rightarrow\{\bv_{\so},\bv_{\st}\}$ and $\bu\Rightarrow\{\bu_{\so},\bu_{\st}\}$ with the restriction $\#\bu_{\so}=\#\bv_{\so}=n$, where $n=0,1,\dots,\min(a,b)$. We recall also that we use the shorthand notation for the products of all the functions and the operators in~\eqref{Phi-expl1}.

An alternative formula for the Bethe vector is
\begin{gather}\label{Phi-expl2}
\BV_{a,b}(\bu;\bv)=\sum K_n(\bv_{\so}|\bu_{\so})\frac{f(\bu_{\so},\bu_{\st}) g(\bv_{\st},\bv_{\so})}
{\lambda_2(\bu_{\st})\lambda_2(\bv)f(\bv,\bu)} \bT_{13}(\bv_{\so}) \bT_{23}(\bv_{\st}) T_{12}(\bu_{\st})\Omega,
\end{gather}
where $K_n$ is the Izergin determinant \eqref{K-def} and the sum is the same as in \eqref{Phi-expl1}.
\end{Definition}

A distinctive feature of the Bethe vectors is that under certain conditions on $\bu$ and $\bv$ (Bethe equations), they become eigenvectors of the transfer matrix. In this case we call them {\it on-shell Bethe vectors}. We will show in Section~\ref{S-OSBV} that the vectors~\eqref{Phi-expl1},~\eqref{Phi-expl2} do possess this property.

It is obvious within the framework of the current approach \cite{KhoPak08} that the action of any mono\-dromy matrix entry $T_{ij}(u)$ on a Bethe vector produces a linear combination of a {\it finite} number of Bethe vectors. This follows from the presentations of the monodromy matrix elements and Bethe vectors in terms of the Cartan--Weyl or current generators of the Yangian double ${\rm DY}(\mathfrak{gl}(2|1))$ and normal ordering of these generators according to certain cyclic ordering~\cite{EKhP,FKPR}. In summary, we can rewrite the action formulas as normal ordering problem for the current generators and then translate the result of this ordering back into f\/inite sum of the Bethe vectors. However, it is not so obvious if we deal with the explicit representations~\eqref{Phi-expl1},~\eqref{Phi-expl2}. Furthermore, in spite of the action of~$T_{ij}(z)$ onto $\BV_{a,b}(\bu;\bv)$ formally can be derived via~\eqref{commTT} and~\eqref{vac}, actually it is a pretty nontrivial problem.

Fortunately, similarly to the $\mathfrak{gl}(3)$ case \cite{BelPRS13a} the $\mathfrak{gl}(2|1)$ Bethe vectors obey recursion relations over the number of the Bethe parameters~\cite{PakRS16a}. The f\/irst recursion has the form
\begin{gather}
T_{12}(z) \BV_{a,b}(\bar u;\bar v)=\lambda_2(z) f(\bv,z)\BV_{a+1,b}(\{\bar u;z\};\bar v) \nonumber\\
\hphantom{T_{12}(z) \BV_{a,b}(\bar u;\bar v)=}{}
+\sum_{j=1}^b g(z,v_j)g(\bv_j,v_j) T_{13}(z) \BV_{a,b-1}(\bar u;\bar v_j).\label{recT12}
\end{gather}
The second recursion reads
\begin{gather}
T_{23}(z)\BV_{a,b}(\bar u;\bar v) =\lambda_2(z)h(\bv,z)f(z,\bu)\BV_{a,b+1}(\bar u;\{\bar v,z\})\nonumber\\
\hphantom{T_{23}(z)\BV_{a,b}(\bar u;\bar v) =}{} +\sum_{j=1}^a g(u_j,z)f(u_j,\bu_j) T_{13}(z) \BV_{a-1,b}(\bar u_j;\bar v).\label{recT23}
\end{gather}
We recall that in these formulas $\bv_j$ and $\bu_j$ respectively mean $\bv\setminus \{v_j\}$ and $\bu\setminus \{u_j\}$. The shorthand notation for the products of the functions~$g$ and~$f$ is also used.

Equations \eqref{recT12} and \eqref{recT23} allow us to built recursively Bethe vectors starting with the simplest cases
\begin{gather}\label{simp-BV}
\BV_{a,0}(\bar u;\varnothing)=\frac{T_{12}(\bu)}{\lambda_{2}(\bu)}\Omega,\qquad
\BV_{0,b}(\varnothing; \bar v)=\frac{\bT_{23}(\bv)}{\lambda_{2}(\bv)}\Omega.
\end{gather}
One can also easily derive the actions of $T_{ij}$ onto either $\BV_{a,0}(\bar u;\varnothing)$ or $\BV_{0,b}(\varnothing; \bar v)$, and then, using induction over $a$ or $b$ obtain the action rule in the general case. This will be our main strategy in the derivation of the action formulas.

\section[Multiple actions of the operators $T_{ij}$ onto Bethe vectors]{Multiple actions of the operators $\boldsymbol{T_{ij}}$ onto Bethe vectors}\label{Mult-action}

The main result of this paper consists of explicit formulas of the multiple actions of the mo\-no\-dromy matrix entries onto Bethe vectors. We show that these actions always reduce to f\/inite linear combinations of Bethe vectors.

Everywhere in this section we assume that $\bu$, $\bv$, and $\bar z$ are three sets of generic complex numbers with cardinalities $\#\bu=a$, $\#\bv=b$, and $\#\bar z=n$, $a,b,n=0,1,\dots$. We also set $\bet=\{\bu,\bar z\}$ and $\bxi=\{\bv,\bar z\}$.

\subsection[Actions of $T_{ij}(\bar z)$ with $i<j$]{Actions of $\boldsymbol{T_{ij}(\bar z)}$ with $\boldsymbol{i<j}$}

\begin{itemize}\itemsep=0pt
\item Multiple action of $T_{13}(z)$:
\begin{gather}\label{A-T13}
\bT_{13}(\bar z)\BV_{a,b}(\bu;\bv)=\lambda_2(\bar z)h(\bv,\bar z)\BV_{a+n,b+n}(\bet;\bxi).
\end{gather}

\item Multiple action of $T_{12}(z)$:
\begin{gather}\label{A-T12}
T_{12}(\bar z)\BV_{a,b}(\bu;\bv)=\lambda_2(\bar z)h(\bxi,\bar z)\sum \frac{g(\bxi_{\st},\bxi_{\so})}{h(\bxi_{\so},\bar z)}\BV_{a+n,b}(\bet;\bxi_{\st}).
\end{gather}
Here the sum is taken over partitions $\bxi\Rightarrow\{\bxi_{\so},\bxi_{\st}\}$ with $\#\bxi_{\so}=n$.

\item Multiple action of $T_{23}(z)$:
\begin{gather}\label{A-T23}
\bT_{23}(\bar z)\BV_{a,b}(\bu;\bv)=(-1)^n\lambda_2(\bar z)h(\bv,\bar z)\sum K_n(\bar z|\bet_{\so}+c)f(\bet_{\so},\bet_{\st})\BV_{a,b+n}(\bet_{\st};\bxi).
\end{gather}
Here the sum is taken over partitions $\bet\Rightarrow\{\bet_{\so},\bet_{\st}\}$ with $\#\bet_{\so}=n$.
\end{itemize}

\subsection[Actions of $T_{ii}(\bar z)$]{Actions of $\boldsymbol{T_{ii}(\bar z)}$}

In formulas \eqref{A-T11}--\eqref{A-T33} the sums are taken over partitions $\bxi\Rightarrow\{\bxi_{\so},\bxi_{\st}\}$ and
$\bet\Rightarrow\{\bet_{\so},\bet_{\st}\}$ with $\#\bxi_{\so}=\#\bet_{\so}=n$.
\begin{itemize}\itemsep=0pt
\item Multiple action of $T_{11}(z)$:
\begin{gather}
T_{11}(\bar z)\BV_{a,b}(\bu;\bv)=(-1)^n\lambda_2(\bar z)h(\bxi,\bar z)\nonumber\\
\hphantom{T_{11}(\bar z)\BV_{a,b}(\bu;\bv)=}{}\times\sum r_1(\bet_{\so})
\frac{f(\bet_{\st},\bet_{\so})g(\bxi_{\st},\bxi_{\so})}{h(\bxi_{\so},\bar z)f(\bxi_{\st},\bet_{\so})}K_n(\bet_{\so}|\bxi_{\so}+c)
\BV_{a,b}(\bet_{\st};\bxi_{\st}).\label{A-T11}
\end{gather}

\item Multiple action of $T_{22}(z)$:
\begin{gather}
T_{22}(\bar z)\BV_{a,b}(\bu;\bv)=(-1)^n\lambda_2(\bar z)h(\bxi,\bar z)\nonumber\\
\hphantom{T_{22}(\bar z)\BV_{a,b}(\bu;\bv)=}{}\times\sum
\frac{f(\bet_{\so},\bet_{\st})g(\bxi_{\st},\bxi_{\so})}{h(\bxi_{\so},\bar z)}K_n(\bar z|\bet_{\so}+c)
\BV_{a,b}(\bet_{\st};\bxi_{\st}).\label{A-T22}
\end{gather}

\item Multiple action of $T_{33}(z)$:
\begin{gather}\label{A-T33}
T_{33}(\bar z)\BV_{a,b}(\bu;\bv)=\lambda_2(\bar z)h(\bxi,\bar z)\sum r_3(\bxi_{\so})
\frac{f(\bet_{\so},\bet_{\st})g(\bxi_{\st},\bxi_{\so})h(\bet_{\so},\bet_{\so})}{h(\bxi_{\so},\bet_{\so})h(\bet_{\so},\bar z)f(\bxi_{\so},\bet_{\st})}
\BV_{a,b}(\bet_{\st};\bxi_{\st}).
\end{gather}
\end{itemize}

\subsection[Actions of $T_{ij}(\bar z)$ with $i>j$]{Actions of $\boldsymbol{T_{ij}(\bar z)}$ with $\boldsymbol{i>j}$}

\begin{itemize}\itemsep=0pt
\item Multiple action of $T_{21}(z)$:
\begin{gather}
T_{21}(\bar z)\BV_{a,b}(\bu;\bv)=\lambda_2(\bar z)h(\bxi,\bar z)\sum r_1(\bet_{\so})
\frac{f(\bet_{\st},\bet_{\so})f(\bet_{\st},\bet_{\sth})f(\bet_{\sth},\bet_{\so})g(\bxi_{\st},\bxi_{\so})}
{h(\bxi_{\so},\bar z)f(\bxi_{\st},\bet_{\so})}\nonumber\\
\hphantom{T_{21}(\bar z)\BV_{a,b}(\bu;\bv)=}{}
\times K_n(\bar z|\bet_{\st}+c)K_n(\bet_{\so}|\bxi_{\so}+c)
\BV_{a-n,b}(\bet_{\sth};\bxi_{\st}).\label{A-T21}
\end{gather}
Here the sum is taken over partitions $\bxi\Rightarrow\{\bxi_{\so},\bxi_{\st}\}$ and
$\bet\Rightarrow\{\bet_{\so},\bet_{\st},\bet_{\sth}\}$ with $\#\bxi_{\so}=\#\bet_{\so}=\#\bet_{\st}=n$.

\item Multiple action of $T_{32}(z)$:
\begin{gather}
\bT_{32}(\bar z)\BV_{a,b}(\bu;\bv)=(-1)^{\frac{n(n-1)}2}\lambda_2(\bar z)h(\bxi,\bar z)\nonumber\\
\qquad{}\times \sum r_3(\bxi_{\so})
\frac{f(\bet_{\so},\bet_{\st}) g(\bxi_{\st},\bxi_{\so})g(\bxi_{\sth},\bxi_{\st})g(\bxi_{\sth},\bxi_{\so})}
{h(\bet_{\so},\bar z)h(\bxi_{\so},\bet_{\so})h(\bxi_{\st},\bar z)f(\bxi_{\so},\bet_{\st})}
 h(\bet_{\so},\bet_{\so}) \BV_{a,b-n}(\bet_{\st};\bxi_{\sth}).\label{A-T32}
\end{gather}
Here the sum is taken over partitions $\bxi\Rightarrow\{\bxi_{\so},\bxi_{\st},\bxi_{\sth}\}$ and
$\bet\Rightarrow\{\bet_{\so},\bet_{\st}\}$ with $\#\bxi_{\so}=\#\bxi_{\st}=\#\bet_{\so}=n$.

\item Multiple action of $T_{31}(z) $:
\begin{gather}
\bT_{31}(\bar z)\BV_{a,b}(\bu;\bv)=(-1)^{\frac{n(n+1)}2}\lambda_2(\bar z)h(\bxi,\bar z)\sum r_3(\bxi_{\so})r_1(\bet_{\st})
\frac{g(\bxi_{\st},\bxi_{\so})g(\bxi_{\sth},\bxi_{\st})g(\bxi_{\sth},\bxi_{\so})}
{h(\bet_{\so},\bar z)h(\bxi_{\so},\bet_{\so})h(\bxi_{\st},\bar z)}\nonumber\\
\qquad{}
\times \frac{f(\bet_{\so},\bet_{\st}) f(\bet_{\so},\bet_{\sth}) f(\bet_{\sth},\bet_{\st}) h(\bet_{\so},\bet_{\so})}{f(\bxi_{\so},\bet_{\st})f(\bxi_{\so},\bet_{\sth})f(\bxi_{\sth},\bet_{\st})}
 K_n(\bet_{\st}|\bxi_{\st}+c) \BV_{a-n,b-n}(\bet_{\sth};\bxi_{\sth}).\label{A-T31}
\end{gather}
Here the sum is taken over partitions $\bxi\Rightarrow\{\bxi_{\so},\bxi_{\st},\bxi_{\sth}\}$ and
$\bet\Rightarrow\{\bet_{\so},\bet_{\st},\bet_{\sth}\}$ with $\#\bxi_{\so}=\#\bxi_{\st}=\#\bet_{\so}=\#\bet_{\st}=n$.
\end{itemize}

The proofs of the multiple action formulas will be given in Sections~\ref{ilessj}--\ref{Induction}.

\section{On-shell Bethe vectors}\label{S-OSBV}

The action formulas \eqref{A-T13}--\eqref{A-T31} are valid for generic complex numbers $\bar z$, $\bu$, and $\bv$. In this section we consider them for on-shell Bethe vector $\BV_{a,b}(\bu;\bv)$, that is when the parameters~$\bu$ and~$\bv$ satisfy a system of Bethe equations (see~\eqref{AEigenS-1}).

In order to f\/ind explicitly the result of the transfer matrix action onto $\BV_{a,b}(\bu;\bv)$ one should set $n=1$ in \eqref{A-T11}--\eqref{A-T33}. Then the subsets $\bet_{\so}$ and $\bxi_{\so}$ consist of one element only. Obviously, there are two essentially dif\/ferent types of partitions of the set $\bet=\{z,\bu\}$:
\begin{alignat}{3}
& \bet_{\so} = z,\qquad && \bet_{\st}=\bu,& \label{part-bet1}\\
& \bet_{\so} = u_j,\qquad && \bet_{\st}=\{z,\bu_j\},\qquad j=1,\dots,a.& \label{part-bet2}
\end{alignat}
Similarly, there are two dif\/ferent types of partitions of the set $\bxi=\{z,\bv\}$:
\begin{alignat}{3}
& \bxi_{\so} = z,\qquad && \bxi_{\st}=\bv,& \label{part-xi1}\\
& \bxi_{\so} = v_k,\qquad && \bxi_{\st}=\{z,\bv_k\},\qquad k=1,\dots,b.&\label{part-xi2}
\end{alignat}
Thus, the action of $\mathcal{T}(z)$ onto $\BV_{a,b}(\bu;\bv)$ can be written in the form
\begin{gather*}
\mathcal{T}(z)\BV_{a,b}(\bu;\bv)=\tau(z|\bu;\bv)\BV_{a,b}(\bu;\bv)+\sum_{j=1}^a \Lambda_j \BV_{a,b}(\{z,\bu_j\};\bv)\nonumber\\
\hphantom{\mathcal{T}(z)\BV_{a,b}(\bu;\bv)=}{} +\sum_{k=1}^b \tilde\Lambda_k \BV_{a,b}(\bu;\{z,\bv_k\})+\sum_{j=1}^a\sum_{k=1}^b M_{jk} \BV_{a,b}(\{z,\bu_j\};\{z,\bv_k\}),
\end{gather*}
where $\tau$, $\Lambda_j$, $\tilde\Lambda_k$, and $M_{jk}$ are numerical coef\/f\/icients. In order to f\/ind $\tau(z|\bu;\bv)$ we substitute the partitions \eqref{part-bet1} and \eqref{part-xi1} into \eqref{A-T11}--\eqref{A-T33}. We obtain
\begin{gather}\label{tau}
\tau(z|\bu,\bv)=\lambda_1(z)f(\bu,z)+\lambda_2(z)f(z,\bu)f(\bv,z)-\lambda_3(z)f(\bv,z),
\end{gather}
where we have used $h(z,z)=1$ and $K_1(z|z+c)=g(z,z+c)=-1$.

In order to f\/ind $\Lambda_j$ we substitute the partitions \eqref{part-bet2} and \eqref{part-xi1} into \eqref{A-T11}--\eqref{A-T33}. We f\/ind
\begin{gather*}
\Lambda_j=\lambda_2(z)h(\bv,z)g(\bv,z)g(z,u_j)\left( r_1(u_j) \frac{f(\bu_j,u_j)}{f(\bv,u_j)}-f(u_j,\bu_j)\right).
\end{gather*}
Similarly, in order to f\/ind $\tilde\Lambda_k$ we substitute the partitions \eqref{part-bet1} and \eqref{part-xi2} into \eqref{A-T11}--\eqref{A-T33}. This gives us
\begin{gather*}
\tilde\Lambda_k=\lambda_2(z)f(z,\bu)g(\bv_k,v_k)h(\bv_k,z)g(z,v_k)\left(1-\frac{ r_3(v_k)}{f(v_k,\bu)}\right).
\end{gather*}
If $\BV_{a,b}(\bu;\bv)$ is an eigenvector of $\mathcal{T}(z)$, then the coef\/f\/icients $\Lambda_j$ and $\tilde\Lambda_k$ must vanish for arbitrary~$z$. Setting $\Lambda_j=0$ for $j=1,\dots, a$ and $\tilde\Lambda_k=0$ for $k=1,\dots,b$ we arrive at a system of equations
\begin{gather}
r_1(u_j) =\frac{f(u_j,\bu_j)}{f(\bu_j,u_j)}f(\bv,u_j),\qquad j=1,\dots,a,\nonumber\\
r_3(v_k) =f(v_k,\bu),\qquad k=1,\dots,b.\label{AEigenS-1}
\end{gather}

Let us check that $M_{jk}=0$ provided the system \eqref{AEigenS-1} is fulf\/illed. Substituting the parti\-tions~\eqref{part-bet2} and~\eqref{part-xi2} into \eqref{A-T11}--\eqref{A-T33} we obtain
\begin{gather*}
M_{jk}=\lambda_2(z)h(\bv_k,z)g(\bv_k,v_k)\Biggl(
\frac{r_1(u_j)f(\bu_j,u_j)}{f(\bv,u_j)}g(v_k,u_j)g(z,v_k)\\
\hphantom{M_{jk}=}{}
+f(u_j,\bu_j)g(u_j,z)\left[g(z,v_k)+\frac{r_3(v_k)}{f(v_k,\bu)}g(v_k,u_j)\right]\Biggr).
\end{gather*}
Substituting here $r_1(u_j)$ and $r_3(v_k)$ from equations \eqref{AEigenS-1}, we immediately f\/ind that $M_{jk}=0$ due to the identity
\begin{gather*}
g(v_k,u_j)g(z,v_k)+g(u_j,z)g(z,v_k)+g(u_j,z)g(v_k,u_j)=0.
\end{gather*}

Thus, the system \eqref{AEigenS-1} can be treated as the system of Bethe equations for the parameters~$\bu$ and~$\bv$. If \eqref{AEigenS-1} holds,
then the corresponding Bethe vector $\BV_{a,b}(\bu;\bv)$ is on-shell, i.e., it is an eigenvector of the transfer matrix $\mathcal{T}(z)$. The eigenvalue of this on-shell vector is given by~\eqref{tau}. At the same time, it is easy to see that the function $\tau(z|\bu,\bv)$ has no poles in the points $z=u_j$, $j=1,\dots,a$, and $z=v_k$, $k=1,\dots,b$ due to the system~\eqref{AEigenS-1}.

\section[Proofs of multiple actions for $T_{ij}$ with $i<j$]{Proofs of multiple actions for $\boldsymbol{T_{ij}}$ with $\boldsymbol{i<j}$} \label{ilessj}

Bethe vectors consist of the elements from the upper triangular part of the monodromy matrix applied to pseudovacuum~$\Omega$~\eqref{Phi-expl1},~\eqref{Phi-expl2}. Then, it is intuitively clear that actions of the elements~$T_{ij}$ with $i<j$ are the simplest. We begin our consideration from the right-upper corner of monodromy matrix and will move along anti-diagonal direction successively proving the action relations.

\subsection[Proof for $T_{13}$]{Proof for $\boldsymbol{T_{13}}$}

For $n=1$ equation \eqref{A-T13} follows directly from the def\/initions of the Bethe vectors. Let us take, for instance,~\eqref{Phi-expl1}
and set there $\bu=\{z,\bu'\}$ and $\bv=\{z,\bv'\}$. Then the product $1/f(\bv,\bu)$ vanishes, as it contains $1/f(z,z)$. This zero, however,
can be compensated if and only if $z\in\bu_{\so}$ and $z\in\bv_{\so}$. Indeed, in this case the product $g(\bv_{\so}, \bu_{\so})$ contains a
singular factor $g(z,z)$. Thus, we should consider only such partitions, for which $z\in\bu_{\so}$ and $z\in\bv_{\so}$. Therefore we should set:
$\bu_{\so}=\{z,\bu'_{\so}\}$ and $\bv_{\so}=\{z,\bv'_{\so}\}$; $\bu_{\st}=\bu'_{\st}$ and $\bv_{\st}=\bv'_{\st}$. Then we obtain
\begin{gather*}
\BV_{a,b}(\{z,\bu'\};\{z,\bv'\})=\sum g(\bv'_{\so},\bu'_{\so}) \frac{ f(\bu'_{\so},\bu'_{\st}) g(\bv'_{\st},\bv'_{\so})h(\bu'_{\so},\bu'_{\so})}
{\lambda_2(\bu')\lambda_2(\bv'_{\st})f(\bv',\bu')} g(\bv'_{\so},z) g(z,\bu'_{\so})\nonumber\\
\hphantom{\BV_{a,b}(\{z,\bu'\};\{z,\bv'\})=}{}
\times \frac{ f(z,\bu'_{\st}) g(\bv'_{\st},z)h(z,\bu'_{\so})h(\bu'_{\so},z)}
{\lambda_2(z)f(\bv',z)f(z,\bu')} \frac{T_{13}(z)}{h(\bu'_{\so},z)}
\bT_{13}(\bu'_{\so}) T_{12}(\bu'_{\st}) \bT_{23}(\bv'_{\st})\Omega.
\end{gather*}
After evident cancellations we arrive at
\begin{gather*}
\BV_{a,b}(\{z,\bu'\};\{z,\bv'\})= \frac{T_{13}(z)}{\lambda_2(z)h(\bv',z)} \BV_{a-1,b-1}(\bu';\bv'),
\end{gather*}
which coincides with \eqref{A-T13} at $n=1$. The same result arises from the analysis of equation \eqref{Phi-expl2}.

Now we use induction over $n$. Assume that \eqref{A-T13} holds for some $n-1$. Then
\begin{gather*}
\bT_{13}(\bar z)\BV_{a,b}(\bu;\bv)=\frac{T_{13}(z_n)\bT_{13}(\bar z_n)}{h(\bar z_n,z_n)} \BV_{a,b}(\bu;\bv)\\
\hphantom{\bT_{13}(\bar z)\BV_{a,b}(\bu;\bv)}{} =\lambda_2(\bar z_n)\frac{h(\bv,\bar z_n)}{h(\bar z_n,z_n)} T_{13}(z_n)\BV_{a+n-1,b+n-1}(\{\bu,\bar z_n\};\{\bv,\bar z_n\})\\
\hphantom{\bT_{13}(\bar z)\BV_{a,b}(\bu;\bv)}{} =\lambda_2(\bar z)h(\bv,\bar z)\BV_{a+n,b+n}(\{\bu,\bar z\};\{\bv,\bar z\}),
\end{gather*}
and thus, \eqref{A-T13} is proved.

Using \eqref{A-T13} one can recast recursions \eqref{recT12} and \eqref{recT23} as follows
\begin{gather}
T_{12}(z) \BV_{a,b}(\bar u;\bar v)=\lambda_2(z) f(\bv,z)\BV_{a+1,b}(\{\bar u,z\};\bar v) \nonumber\\
\hphantom{T_{12}(z) \BV_{a,b}(\bar u;\bar v)=}{} +\lambda_2(z)\sum_{j=1}^b g(z,v_j)g(\bv_j,v_j)h(\bv_j,z)
 \BV_{a+1,b}(\{\bar u,z\};\{\bar v_j,z\}),\label{recT12-n}
\end{gather}
and
\begin{gather*}
T_{23}(z)\BV_{a,b}(\bar u;\bar v) =\lambda_2(z)h(\bv,z)\biggl(f(z,\bu)\BV_{a,b+1}(\bar u;\{\bar v,z\})\\
\hphantom{T_{23}(z)\BV_{a,b}(\bar u;\bar v) =}{} +\sum_{j=1}^a g(u_j,z)f(u_j,\bu_j) \BV_{a,b+1}(\{\bar u_j,z\};\{\bar v,z\})\biggr).
\end{gather*}
One can easily recognize in these equations the actions~\eqref{A-T12} and~\eqref{A-T23} for $n=1$. Then one should use induction over~$n$.

\subsection[Proof for $T_{12}$]{Proof for $\boldsymbol{T_{12}}$}

Assume that \eqref{A-T12} holds for some $n-1$. Then
\begin{gather*}
T_{12}(\bar z)\BV_{a,b}(\bu;\bv)=T_{12}(z_n)\lambda_2(\bar z_n)h(\bxi,\bar z_n)
\sum \frac{g(\bxi_{\st},\bxi_{\so})}{h(\bxi_{\so},\bar z_n)}\BV_{a+n-1,b}(\bet;\bxi_{\st}).
\end{gather*}
Here $\bet=\{\bar z_n,\bu\}$, $\bxi=\{\bar z_n,\bv\}$, and the sum runs through the partitions $\bxi\Rightarrow \{\bxi_{\so},\bxi_{\st}\}$ with $\#\bxi_{\so}=n-1$. Acting with $T_{12}(z_n)$ we obtain
\begin{gather*}
T_{12}(\bar z)\BV_{a,b}(\bu;\bv)=\lambda_2(\bar z)\frac{h(\bxi,\bar z_n)}{h(z_n,\bar z_n)}
\sum \frac{g(\bxi_{\st},\bxi_{\so})}{h(\bxi_{\so},\bar z_n)}h(\bxi_{\st},z_n)
\frac{g(\bxi_{\rm ii},\bxi_{\rm i})}{h(\bxi_{\rm i},z_n)}
\BV_{a+n,b}(\bet;\bxi_{\rm ii}).
\end{gather*}
Here already $\bet=\{\bar z,\bu\}$ and $\bxi=\{\bar z,\bv\}$. The sum f\/irst is taken over partitions $\{\bar z_n,\bv\}\Rightarrow\{\bxi_{\so},\bxi_{\st}\}$ with $\#\bxi_{\so}=n-1$, and then over partitions $\{z_n,\bxi_{\st}\}\Rightarrow\{\bxi_{\rm i},\bxi_{\rm ii}\}$ with $\#\bxi_{\rm i}=1$. One can say that the sum is taken over partitions $\{\bar z,\bv\}=\bxi\Rightarrow\{\bxi_{\so},\bxi_{\rm i},\bxi_{\rm ii}\}$ with restrictions $z_n\notin \bxi_{\so}$, $\#\bxi_{\so}=n-1$, and $\#\bxi_{\rm i}=1$. Presenting $\bxi_{\st}$ as $\bxi_{\st}=\{\bxi_{\rm i},\bxi_{\rm ii}\}\setminus \{z_n\}$ we obtain
\begin{gather*}
g(\bxi_{\st},\bxi_{\so})=\frac{g(\bxi_{\rm i},\bxi_{\so})g(\bxi_{\rm ii},\bxi_{\so})}{g(z_n,\bxi_{\so})},\qquad
h(\bxi_{\st},z_n) =h(\bxi_{\rm i}, z_n)h(\bxi_{\rm ii}, z_n),
\end{gather*}
and hence,
\begin{gather}\label{A-T12n-3}
T_{12}(\bar z)\BV_{a,b}(\bu;\bv)=\lambda_2(\bar z)\frac{h(\bxi,\bar z_n)}{h(z_n,\bar z_n)}
\sum \frac{g(\bxi_{\rm i},\bxi_{\so})g(\bxi_{\rm ii},\bxi_{\so})g(\bxi_{\rm ii},\bxi_{\rm i})}
{g(z_n,\bxi_{\so})h(\bxi_{\so},\bar z_n)}h(\bxi_{\rm ii},z_n)
 \BV_{a+n,b}(\bet;\bxi_{\rm ii}).
\end{gather}
Observe that the condition $z_n\notin \bxi_{\so}$ is ensured by the product $g(z_n,\bxi_{\so})$ in the denominator. Hen\-ce, we can say that the sum is taken over partitions $\bxi\Rightarrow\{\bxi_{\so},\bxi_{\rm i},\bxi_{\rm ii}\}$ with the restrictions on the cardinalities of the subsets only. Setting $\{\bxi_{\so},\bxi_{\rm i}\}=\bxi_{0}$ we recast \eqref{A-T12n-3} as follows
\begin{gather*}
T_{12}(\bar z)\BV_{a,b}(\bu;\bv)=\lambda_2(\bar z)\frac{h(\bxi,\bar z)}{h(z_n,\bar z_n)}
\sum \frac{g(\bxi_{\rm i},\bxi_{\so})g(z_n,\bxi_{\rm i})h(\bxi_{\rm i},\bar z_n)}
{g(z_n,\bxi_{0})h(\bxi_{0},\bar z)} g(\bxi_{\rm ii},\bxi_{0})
 \BV_{a+n,b}(\bet;\bxi_{\rm ii}).
\end{gather*}
The sum over partitions $\bxi_{0}\Rightarrow \{\bxi_{\so},\bxi_{\rm i}\}$ can be computed via Lemma~\ref{main-ident-C}
\begin{gather}\label{CI}
\sum_{\bxi_{0}\Rightarrow \{\bxi_{\so},\bxi_{\rm i}\}}
 g(\bxi_{\rm i},\bxi_{\so})g(z_n,\bxi_{\rm i})h(\bxi_{\rm i},\bar z_n)=g(z_n,\bxi_{0})h(z_n,\bar z_n),
\end{gather}
where we took into account that $\#\bxi_{0}=n$. Thus, we arrive at
\begin{gather*}
T_{12}(\bar z)\BV_{a,b}(\bu;\bv)=\lambda_2(\bar z)h(\bxi,\bar z) \sum \frac{ g(\bxi_{\rm ii},\bxi_{0})}{h(\bxi_{0},\bar z)}
 \BV_{a+n,b}(\bet;\bxi_{\rm ii}),
\end{gather*}
 which coincides with \eqref{A-T12} up to a relabeling of the subsets.

\subsection[Proof for $T_{23}$]{Proof for $\boldsymbol{T_{23}}$}

Assume that \eqref{A-T23} holds for some $n-1$. Let $\#\bar z=n$. Then we have
\begin{gather*}
\bT_{23}(\bar z)\BV_{a,b}(\bu;\bv)=\frac{T_{23}(z_n)\bT_{23}(\bar z_n)}{h(\bar z_n,z_n)} \BV_{a,b}(\bu;\bv)\\
\qquad{} =(-1)^{n-1}\lambda_2(\bar z_n)\frac{h(\bv,\bar z_n)}{h(\bar z_n,z_n)}
\sum K_{n-1}(\bar z_n|\bet_{\so}+c)f(\bet_{\so},\bet_{\st})T_{23}(z_n)\BV_{a,b+n-1}(\bet_{\st};\bxi).
\end{gather*}
Here $\bet=\{\bar z_n,\bu\}$, $\bxi=\{\bar z_n,\bv\}$, and the sum runs through the partitions $\bet\Rightarrow \{\bet_{\so},\bet_{\st}\}$ with $\#\bet_{\so}=n-1$. Acting with $T_{23}(z_n)$ we obtain
\begin{gather*}
\bT_{23}(\bar z)\BV_{a,b}(\bu;\bv)=(-1)^{n}\lambda_2(\bar z)\frac{h(\bv,\bar z_n)}{h(\bar z_n,z_n)}\\
\qquad{} \times \sum K_{n-1}(\bar z_n|\bet_{\so}+c)f(\bet_{\so},\bet_{\st})h(\bxi,z_n)
K_{1}(z_n|\bet_{\rm i}+c)f(\bet_{\rm i},\bet_{\rm ii})\BV_{a,b+n}(\bet_{\rm ii};\bxi).
\end{gather*}
Here already $\bet=\{\bar z,\bu\}$, $\bxi=\{\bar z,\bv\}$, and the sum is taken f\/irst over partitions $\{\bar z_n,\bu\}\Rightarrow \{\bet_{\so},\bet_{\st}\}$ with $\#\bet_{\so}=n-1$, and then over partitions $\{z_n,\bet_{\st}\}\Rightarrow \{\bet_{\rm i},\bet_{\rm ii}\}$ with $\#\bet_{\rm i}=1$. Substituting here $\bet_{\st}=\{\bet_{\rm i},\bet_{\rm ii}\}\setminus \{z_n\}$ we f\/ind
\begin{gather*}
\bT_{23}(\bar z)\BV_{a,b}(\bu;\bv)=(-1)^{n}\lambda_2(\bar z)h(\bv,\bar z)\\
\qquad{}
\times \sum K_{n-1}(\bar z_n|\bet_{\so}+c)K_{1}(z_n|\bet_{\rm i}+c)
\frac{f(\bet_{\so},\bet_{\rm i})f(\bet_{\so},\bet_{\rm ii})f(\bet_{\rm i},\bet_{\rm ii})}{f(\bet_{\so},z_n)}
\BV_{a,b+n}(\bet_{\rm ii};\bxi).
\end{gather*}
Setting $\bet_0=\{\bet_{\so},\bet_{\rm i}\}$ and using $K_{1}(z_n|\bet_{\rm i}+c)=-K_{1}(\bet_{\rm i}|z_n)/f(\bet_{\rm i},z_n)$ we obtain
\begin{gather*}
\bT_{23}(\bar z)\BV_{a,b}(\bu;\bv)=(-1)^{n-1}\lambda_2(\bar z)h(\bv,\bar z)\\
\hphantom{\bT_{23}(\bar z)\BV_{a,b}(\bu;\bv)=}{}
\times \sum K_{n-1}(\bar z_n|\bet_{\so}+c)K_{1}(\bet_{\rm i}|z_n)f(\bet_{\so},\bet_{\rm i})
\frac{f(\bet_{0},\bet_{\rm ii})}{f(\bet_{0},z_n)}
\BV_{a,b+n}(\bet_{\rm ii};\bxi).
\end{gather*}
Now we can compute the sum over partitions $\bet_{0}\Rightarrow\{\bet_{\so},\bet_{\rm i}\}$ via \eqref{Sym-Part-old1}
\begin{gather}\label{ML}
\sum_{\bet_{0}\Rightarrow\{\bet_{\so},\bet_{\rm i}\}} K_{n-1}(\bar z_n|\bet_{\so}+c)K_{1}(\bet_{\rm i}|z_n)f(\bet_{\so},\bet_{\rm i})=
-f(\bet_{0},z_n) K_{n}(\bar z|\bet_{0}+c),
\end{gather}
which gives us
\begin{gather*}
\bT_{23}(\bar z)\BV_{a,b}(\bu;\bv)=(-1)^{n}\lambda_2(\bar z)h(\bv,\bar z)
\sum K_{n}(\bar z|\bet_{0}+c)f(\bet_{0},\bet_{\rm ii})\BV_{a,b+n}(\bet_{\rm ii};\bxi).
\end{gather*}
This is exactly \eqref{A-T23} up to the labeling of the subsets.

\section[Proof of the multiple action of the operator $T_{22}$]{Proof of the multiple action of the operator $\boldsymbol{T_{22}}$}

The proofs for the actions \eqref{A-T11}--\eqref{A-T31} are much more involved than the ones considered in the previous section. Fortunately,
they all are quite similar. Therefore, we only detail one as a~typical example, the other actions being proven in the same manner. We focus on the opera\-tor~$T_{22}(u)$.

The strategy of the proof is the following. First, we prove equation \eqref{A-T22} for $a=\#\bu=0$ and $n=\#\bar z=1$. This can be done either via the standard consideration of the algebraic Bethe ansatz or using induction over $b=\#\bv$. In both cases we use~\eqref{simp-BV} and the relation
\begin{gather}\label{ComRel1}
T_{22}(u)T_{23}(v)=f(v,u)T_{23}(v)T_{22}(u)+g(u,v) T_{23}(u)T_{22}(v),
\end{gather}
that follows from \eqref{commTT}.

The next step of the proof is an induction over~$a$. We assume that \eqref{A-T22} is valid for $n=1$ and some~$a$ and use recursion~\eqref{recT12}.
Hereby, we use some of commutation relations \eqref{commTT}
\begin{gather}
T_{22}(u)T_{12}(v)=f(u,v)T_{12}(v)T_{22}(u)+g(v,u) T_{12}(u)T_{22}(v),\label{ComRel2}\\
[T_{22}(u),T_{13}(v)]=g(u,v)\bigl(T_{12}(v)T_{23}(u)-T_{12}(u)T_{23}(v)\bigr).\label{ComRel3}
\end{gather}
Finally, when equation \eqref{A-T22} is proved for $n=1$ and arbitrary $a$ and $b$ we use induction over $n$.

\begin{Remark}
We begin the proof with the case $n=1$, $a=0$, and arbitrary $b$. However, one could also begin with the case $n=1$, $b=0$, and arbitrary~$a$.
For the action of the operator $T_{22}(z)$ this is a matter of choice. For other operators these two starting cases could be essentially dif\/ferent. For instance, one can easily see that $T_{21}(z)\BV_{0,b}(\varnothing,\bv)=0$ for arbitrary $b$. On the other hand, the action $T_{21}(z) \BV_{a,0}(\bu,\varnothing)$ is highly nontrivial, although it is clear that it should coincide with the analogous action in the models with $\mathfrak{gl}(3)$-invariant $R$-matrix. Obviously, in this case it is better to begin the proof with the vector $\BV_{0,b}(\varnothing,\bv)$.
\end{Remark}

\subsection[Action of $T_{22}(z)$ at $a=0$ and $z=1$]{Action of $\boldsymbol{T_{22}(z)}$ at $\boldsymbol{a=0}$ and $\boldsymbol{z=1}$}

In the particular case $a=0$ and $n=1$ equation \eqref{A-T22} turns into
\begin{gather}\label{A1-T220}
T_{22}(z)\BV_{0,b}(\varnothing;\bv)=\lambda_2(z)h(\bv,z)\sum \frac{g(\bxi_{\st},\bxi_{\so})}{h(\bxi_{\so}, z)} \BV_{0,b}(\varnothing;\bxi_{\st}).
\end{gather}
The sum is taken over partitions $\{z,\bv\}=\bxi\Rightarrow\{\bxi_{\so},\bxi_{\st}\}$ with $\#\bxi_{\so}=1$. We prove this action using the standard scheme of the algebraic Bethe ansatz. The vector $\BV_{0,b}(\varnothing;\bv)$ is given by the second equation~\eqref{simp-BV}. Thus, we should move the operator $T_{22}(z)$ to the right through the product of the operators $T_{23}(v_j)$. Using~\eqref{ComRel1} we easily f\/ind
\begin{gather*}
T_{22}(z)\BV_{0,b}(\varnothing;\bv)=\Lambda \BV_{0,b}(\varnothing;\bv)+\sum_{j=1}^b\Lambda_j \BV_{0,b}(\varnothing;\{\bv_j,z)),
\end{gather*}
where $\Lambda$ and $\Lambda_j$ are some coef\/f\/icients to be determined. Obviously, in order to obtain the coef\/f\/icient of
$\BV_{0,b}(\varnothing;\bv)$ one should use only the f\/irst term in the r.h.s.\ of \eqref{ComRel1}. From this we immediately f\/ind
\begin{gather*}
\Lambda=\lambda_2(z)f(\bv,z).
\end{gather*}

Then, due to the symmetry of $\bT_{23}(\bv)$ over $\bv$ it is enough to f\/ind $\Lambda_1$ only. Permuting $T_{22}(z)$ with $T_{23}(v_1)$ we should use the second term in the r.h.s.\ of \eqref{ComRel1}. We have
\begin{gather*}
T_{22}(z)\frac{\bT_{23}(\bv)}{\lambda_{2}(\bv)}\Omega=T_{22}(z)\frac{T_{23}(v_1)\bT_{23}(\bv_1)}{\lambda_{2}(\bv)h(\bv_1,v_1)}\Omega=
g(z,v_1)T_{23}(z)\frac{T_{22}(v_1)\bT_{23}(\bv_1)}{\lambda_{2}(\bv)h(\bv_1,v_1)}\Omega+UWT,
\end{gather*}
where $UWT$ means {\it unwanted terms}, i.e., the terms that cannot give a contribution to the coef\/f\/icient~$\Lambda_1$. Now, moving $T_{22}(v_1)$ through the product $\bT_{23}(\bv)$ we should use only the f\/irst term in the r.h.s.\ of~\eqref{ComRel1}, which gives us
\begin{gather*}
T_{22}(z)\frac{\bT_{23}(\bv)}{\lambda_{2}(\bv)}\Omega= g(z,v_1)g(\bv_1,v_1)T_{23}(z)\frac{\bT_{23}(\bv_1)}{\lambda_{2}(\bv)}\lambda_{2}(v_1)\Omega+UWT,
\end{gather*}
where we used $g(\bv_1,v_1)=f(\bv_1,v_1)/h(\bv_1,v_1)$. It remains to combine $T_{23}(z)$ and $\bT_{23}(\bv_1)$ into $\bT_{23}(\{z,\bv_1\})$ and we arrive at
\begin{gather*}
T_{22}(z)\frac{\bT_{23}(\bv)}{\lambda_{2}(\bv)}\Omega= \lambda_2(z)g(z,v_1)g(\bv_1,v_1)h(\bv_1,z)\frac{\bT_{23}(\{z,\bv_1\})}{\lambda_{2}(\bv_1)\lambda_2(z)}\Omega+UWT,
\end{gather*}
leading to
\begin{gather*}
\Lambda_1=\lambda_2(z)g(z,v_1)g(\bv_1,v_1)h(\bv_1,z).
\end{gather*}
Thus, we eventually obtain
\begin{gather}
T_{22}(z)\BV_{0,b}(\varnothing;\bv)=\lambda_2(z)f(\bv,z)\BV_{0,b}(\varnothing;\bv)\nonumber\\
\hphantom{T_{22}(z)\BV_{0,b}(\varnothing;\bv)=}{} + \lambda_2(z)
\sum_{j=1}^bg(z,v_j)g(\bv_j,v_j)h(\bv_j,z)\BV_{0,b}(\varnothing;\{\bv_j,z\}).\label{AT22-0b}
\end{gather}
It is easy to see that this formula coincides with \eqref{A1-T220}. Indeed the f\/irst term in \eqref{AT22-0b} corresponds to the partition $\bxi_{\so}=z$ and $\bxi_{\st}=\bv$ in \eqref{A1-T220}. The other terms arise in the case of the partitions $\bxi_{\so}=v_j$, $j=1,\dots,b$, and $\bxi_{\st}=\{z,\bv_j\}$. Thus, action~\eqref{A1-T220} is proved.

\subsection[Induction over $a$]{Induction over $\boldsymbol{a}$}

For $n=1$ equation \eqref{A-T22} takes the form
\begin{gather}\label{A1-T22}
T_{22}(z)\BV_{a,b}(\bu;\bv)=\lambda_2(z)h(\bv,z)\sum
\frac{f(\bet_{\so},\bet_{\st})g(\bxi_{\st},\bxi_{\so})}{h(\bet_{\so},z)h(\bxi_{\so}, z)}\BV_{a,b}(\bet_{\st};\bxi_{\st}).
\end{gather}
The sum is taken over partitions $\bxi\Rightarrow\{\bxi_{\so},\bxi_{\st}\}$ and $\bet\Rightarrow\{\bet_{\so},\bet_{\st}\}$ with $\#\bxi_{\so}=\#\bet_{\so}=1$. We assume that~\eqref{A1-T22} is valid for some $a\ge 0$ and $b$ arbitrary. Then, due to recursion~\eqref{recT12}
we have
\begin{gather}
T_{22}(z_1)\BV_{a+1,b}(\{\bar u;z_2\};\bar v)=T_{22}(z_1)\frac{T_{12}(z_2) \BV_{a,b}(\bar u;\bar v)}{\lambda_2(z_2) f(\bv,z_2)} \nonumber\\
\hphantom{T_{22}(z_1)\BV_{a+1,b}(\{\bar u;z_2\};\bar v)=}{}
-T_{22}(z_1)\sum_{j=1}^b \frac{g(z_2,v_j)g(\bv_j,v_j)
 T_{13}(z_2) \BV_{a,b-1}(\bar u;\bar v_j)}{\lambda_2(z_2) f(\bv,z_2)}.\label{recT12a}
\end{gather}
We see that in order to compute the action of $T_{22}(z_1)$ onto $\BV_{a+1,b}(\{z_2,\bu\};\bv)$ we should calculate the successive actions of the operators $T_{22}(z_1)T_{12}(z_2)$ and $T_{22}(z_1)T_{13}(z_2)$. This can be done via~\eqref{ComRel2} and~\eqref{ComRel3}
\begin{gather}
T_{22}(z_1)T_{12}(z_2)\BV_{a,b}(\bar u;\bar v)\nonumber\\
\qquad{} =\bigl(f(z_1,z_2)T_{12}(z_2)T_{22}(z_1)+g(z_2,z_1) T_{12}(z_1)T_{22}(z_2)\bigr)
\BV_{a,b}(\bar u;\bar v),\label{ComRel2a}\\
T_{22}(z_1)T_{13}(z_2)\BV_{a,b-1}(\bar u;\bar v_j)= T_{13}(z_2)T_{22}(z_1)\BV_{a,b-1}(\bar u;\bar v_j) \nonumber\\
\qquad{} + g(z_1,z_2)\bigl(T_{12}(z_2)T_{23}(z_1) -T_{12}(z_1)T_{23}(z_2)\bigr)\BV_{a,b-1}(\bar u;\bar v_j).\label{ComRel3a}
\end{gather}
Thus, we have reduced the action $T_{22}(z_1)\BV_{a+1,b}(\{\bar u;z_2\};\bar v)$ to the calculation of several successive actions. In all of them
the operator $T_{22}$ acts either on $\BV_{a,b}(\bar u;\bar v)$ or on $\BV_{a,b-1}(\bar u;\bar v_j)$, which are known due to the induction assumption. The actions of other operators $T_{ij}$ with $i<j$ are already known for $a$ and $b$ arbitrary.

\subsubsection[Successive action of $T_{12}$ and $T_{23}$]{Successive action of $\boldsymbol{T_{12}}$ and $\boldsymbol{T_{23}}$}

We begin our calculation with the successive action of the operators $T_{12}$ and $T_{23}$. Using \eqref{A-T23} we have
\begin{gather*}
T_{12}(z_2)T_{23}(z_1)\BV_{a,b}(\bu;\bv)=\lambda_2(z_1)h(\bv,z_1)\sum \frac{f(\bet_{\so},\bet_{\st})}
{h(\bet_{\so},z_1)} T_{12}(z_2)\BV_{a,b+1}(\bet_{\st};\bxi).
\end{gather*}
Here $\bet=\{z_1,\bu\}$ and $\bxi=\{z_1,\bv\}$. The sum is taken over partitions $\bet\Rightarrow\{\bet_{\so},\bet_{\st}\}$
with $\#\bet_{\so}=1$. Then we use \eqref{A-T12} and f\/ind
\begin{gather}
T_{12}(z_2)T_{23}(z_1)\BV_{a,b}(\bu;\bv)\nonumber\\
\qquad{} =\lambda_2(\bar z)h(\bv,z_1)h(\bxi,z_2)\sum \frac{f(\bet_{\so},\bet_{\st})}
{h(\bet_{\so},z_1)}\frac{g(\bxi_{\st},\bxi_{\so})}
{h(\bxi_{\so},z_2)}\BV_{a+1,b+1}(\{\bet_{\st},z_2\};\bxi_{\st}).\label{T12T231-2}
\end{gather}
Here already $\bxi=\{\bar z,\bv\}$, however we still have $\bet=\{z_1,\bu\}$. Replacing $\{\bet_{\st},z_2\}$ by $\bet_{\st}$ we recast~\eqref{T12T231-2} in the form
\begin{gather*}
T_{12}(z_2)T_{23}(z_1)\BV_{a,b}(\bu;\bv)\\
\qquad{} =\lambda_2(\bar z)h(\bv,\bar z)h(z_1,z_2)\sum
\frac{f(\bet_{\so},\bet_{\st})g(\bxi_{\st},\bxi_{\so})}
{f(\bet_{\so},z_2)h(\bet_{\so},z_1)h(\bxi_{\so},z_2)}\BV_{a+1,b+1}(\bet_{\st};\bxi_{\st}).
\end{gather*}
Here $\bet=\{\bar z,\bu\}$, and the sum is taken over partitions $\bet\Rightarrow\{\bet_{\so},\bet_{\st}\}$ and $\bxi\Rightarrow\{\bxi_{\so},\bxi_{\st}\}$ with $\#\bet_{\so}=\#\bxi_{\so}=1$. Note that the condition $z_2\notin \bet_{\so}$ is ensured automatically. Indeed, if $z_2\in \bet_{\so}$, then $1/f(\bet_{\so},z_2)=0$.

Replacing here $z_1\leftrightarrow z_2$ we obtain
\begin{gather*}
T_{12}(z_1)T_{23}(z_2)\BV_{a,b}(\bu;\bv)\\
\qquad{} =\lambda_2(\bar z)h(\bv,\bar z)h(z_2,z_1)\sum \frac{f(\bet_{\so},\bet_{\st})g(\bxi_{\st},\bxi_{\so})}
{f(\bet_{\so},z_1)h(\bet_{\so},z_2)h(\bxi_{\so},z_1)}\BV_{a+1,b+1}(\bet_{\st};\bxi_{\st}).
\end{gather*}
Thus, we f\/ind
\begin{gather}
g(z_1,z_2)\bigl(T_{12}(z_2)T_{23}(z_1)-T_{12}(z_1)T_{23}(z_2)\bigr)\BV_{a,b}(\bu;\bv)\nonumber\\
\qquad{}=\lambda_2(\bar z)h(\bv,\bar z)\sum
\frac{f(\bet_{\so},\bet_{\st})g(\bxi_{\st},\bxi_{\so})}
{h(\bet_{\so},\bar z)}\BV_{a+1,b+1}(\bet_{\st};\bxi_{\st})\nonumber\\
\qquad\quad{}\times\left\{\frac{f(z_1,z_2)}{g(\bet_{\so},z_2)h(\bxi_{\so},z_2)}+\frac{f(z_2,z_1)}{g(\bet_{\so},z_1)h(\bxi_{\so},z_1)}\right\}.\label{T12T231-4}
\end{gather}

\subsubsection[Successive action of $T_{13}$ and $T_{22}$]{Successive action of $\boldsymbol{T_{13}}$ and $\boldsymbol{T_{22}}$}

Combining the actions \eqref{A1-T22} and \eqref{A-T13} we obtain
\begin{gather*}
T_{13}(z_2)T_{22}(z_1)\BV_{a,b}(\bu;\bv)\\
\qquad{} =\lambda_2(\bar z)h(\bv,z_1)\sum
\frac{f(\bet_{\so},\bet_{\st})g(\bxi_{\st},\bxi_{\so})h(\bxi_{\st}, z_2)}{h(\bet_{\so},z_1)h(\bxi_{\so}, z_1)} \BV_{a+1,b+1}(\{\bet_{\st},z_2\};\{\bxi_{\st},z_2\}).
\end{gather*}
Here $\bet=\{\bu,z_1\}$ and $\bxi=\{\bv,z_1\}$. Replacing $\{\bet_{\st},z_2\}$ with $\bet_{\st}$ and $\{\bxi_{\st},z_2\}$ with $\bxi_{\st}$
we arrive at
\begin{gather}
T_{13}(z_2)T_{22}(z_1)\BV_{a,b}(\bu;\bv)=\lambda_2(\bar z)h(\bv,\bar z)h(z_1,z_2)\sum
\frac{f(\bet_{\so},\bet_{\st})g(\bxi_{\st},\bxi_{\so})\BV_{a+1,b+1}(\bet_{\st};\bxi_{\st})}{f(\bet_{\so},z_2)g(z_2,\bxi_{\so})
h(\bet_{\so},z_1)h(\bxi_{\so}, \bar z)}.\!\!\!\label{A2-T13T22}
\end{gather}
Here already $\bet=\{\bu,\bar z\}$ and $\bxi=\{\bv,\bar z\}$. The sum is taken over partitions $\bet\Rightarrow\{\bet_{\so},\bet_{\st}\}$ and $\bxi\Rightarrow\{\bxi_{\so},\bxi_{\st}\}$ with $\#\bet_{\so}=\#\bxi_{\so}=1$.

Now we are able to calculate the successive action $T_{22}(z_1)T_{13}(z_2)$ on $\BV_{a,b}(\bu;\bv)$. Indeed, due to \eqref{ComRel3a} this
successive action is given by a combination of \eqref{T12T231-4} and \eqref{A2-T13T22}. A straightforward calculation leads us to the following representation:
\begin{gather*}
T_{22}(z_1)T_{13}(z_2)\BV_{a,b}(\bu;\bv)=\lambda_2(\bar z)h(\bv,\bar z)h(z_2,z_1)\sum
\frac{f(\bet_{\so},\bet_{\st})g(\bxi_{\st},\bxi_{\so})}{h(\bet_{\so},z_1)h(\bxi_{\so}, z_1)}
\BV_{a+1,b+1}(\bet_{\st};\bxi_{\st}).
\end{gather*}

\begin{Remark}\label{rmk:T22T13}
Taking into account \eqref{A-T13} we conclude that if the action \eqref{A1-T22} is valid on the vector $\BV_{a,b}(\bu;\bv)$, then it is also
valid on vectors of the special type $\BV_{a+1,b+1}(\bu';\bv')$, if $\bu'\cap\bv'\ne\varnothing$.
\end{Remark}

\subsubsection[Successive action of $T_{12}$ and $T_{22}$]{Successive action of $\boldsymbol{T_{12}}$ and $\boldsymbol{T_{22}}$}

Using \eqref{A1-T22} we obtain
\begin{gather*}
T_{12}(z_2)T_{22}(z_1)\BV_{a,b}(\bu;\bv)=\lambda_2(z_1)h(\bv,z_1)\sum
\frac{f(\bet_{\so},\bet_{\st})g(\bxi_{\st},\bxi_{\so})}{h(\bet_{\so},z_1)h(\bxi_{\so}, z_1)}
 T_{12}(z_2)\BV_{a,b}(\bet_{\st};\bxi_{\st}).
\end{gather*}
Here $\bet=\{\bu,z_1\}$ and $\bxi=\{\bv,z_1\}$. Applying \eqref{A-T12} to this formula we f\/ind
\begin{gather*}
T_{12}(z_2)T_{22}(z_1)\BV_{a,b}(\bu;\bv)\\
\qquad{} =\lambda_2(\bar z)h(\bv,z_1)\sum
\frac{f(\bet_{\so},\bet_{\st})g(\bxi_{\st},\bxi_{\so})}{h(\bet_{\so},z_1)h(\bxi_{\so}, z_1)}
 h(\bxi_{\st}, z_2) \frac{g(\bxi_{\rm ii},\bxi_{\rm i})}{h(\bxi_{\rm i}, z_2)} \BV_{a+1,b}(\{\bet_{\st},z_2\};\bxi_{\rm ii}).
\end{gather*}
Here we f\/irst have partitions $\bet=\{\bu,z_1\}\Rightarrow\{\bet_{\so},\bet_{\st}\}$ and $\bxi=\{\bv,z_1\}\Rightarrow \{\bxi_{\so},\bxi_{\st}\}$. Then we combine $\bxi_{\st}$ with $z_2$ and divide this set into new subsets $\{\bxi_{\st},z_2\}\Rightarrow\{\bxi_{\rm i},\bxi_{\rm ii}\}$. The restrictions are: $\#\bxi_{\rm i}= \#\bxi_{\so}=\#\bet_{\so}=1$, $z_2\notin \bet_{\so}$, and $z_2\notin \bxi_{\so}$. As we already did before, we replace $\{\bet_{\st},z_2\}$ with $\bet_{\st}$ and use $\bxi_{\st}=\{\bxi_{\rm i},\bxi_{\rm ii}\}\setminus \{z_2\}$. Then
\begin{gather}
T_{12}(z_2)T_{22}(z_1)\BV_{a,b}(\bu;\bv)=\lambda_2(\bar z)h(\bv,\bar z)h(z_1,z_2)\nonumber\\
\qquad{} \times \sum
\frac{f(\bet_{\so},\bet_{\st})g(\bxi_{\rm i},\bxi_{\so})g(\bxi_{\rm ii},\bxi_{\so})g(\bxi_{\rm ii},\bxi_{\rm i})}
{f(\bet_{\so},z_2)h(\bet_{\so},z_1)h(\bxi_{\so}, z_1)g(z_2,\bxi_{\so}) h(\bxi_{\so}, z_2) h(\bxi_{\rm i}, z_2)}
\BV_{a+1,b}(\bet_{\st};\bxi_{\rm ii}).\label{A3-T12T22}
\end{gather}
Setting $\bxi_0=\{\bxi_{\rm i},\bxi_{\so}\}$ we recast \eqref{A3-T12T22} as follows
\begin{gather*}
T_{12}(z_2)T_{22}(z_1)\BV_{a,b}(\bu;\bv)=\lambda_2(\bar z)h(\bv,\bar z)h(z_1,z_2)\\
\qquad {} \times \sum \frac{f(\bet_{\so},\bet_{\st})g(\bxi_{\rm i},\bxi_{\so})g(\bxi_{\rm ii},\bxi_{0})}
{g(\bet_{\so},z_2)h(\bet_{\so},\bar z)h(\bxi_{\so}, z_1)g(z_2,\bxi_{\so}) h(\bxi_{0}, z_2)}
\BV_{a+1,b}(\bet_{\st};\bxi_{\rm ii}).
\end{gather*}
The sum over partitions of the set $\bxi=\{z_1,z_2,\bv\}$ is organized as follows: f\/irst, we have partitions $\bxi\Rightarrow\{\bxi_{\rm ii},\bxi_{0}\}$; second we divide $\bxi_{0}\Rightarrow\{\bxi_{\rm i},\bxi_{\so}\}$. The latter sum consists of two terms and can be computed straightforwardly. This leads us to
\begin{gather*}
T_{12}(z_2)T_{22}(z_1)\BV_{a,b}(\bu;\bv)=\lambda_2(\bar z)h(\bv,\bar z)h(\bar z,\bar z) \sum
\frac{f(\bet_{\so},\bet_{\st})g(\bxi_{\rm ii},\bxi_{0})}
{g(\bet_{\so},z_2)h(\bet_{\so},\bar z)h(\bxi_{0}, \bar z)} \BV_{a+1,b}(\bet_{\st};\bxi_{\rm ii}),
\end{gather*}
and relabeling $\bxi_{0}\to \bxi_{\so}$, $\bxi_{\rm ii}\to \bxi_{\st}$ we f\/inally obtain
\begin{gather}
T_{12}(z_2)T_{22}(z_1)\BV_{a,b}(\bu;\bv)\nonumber\\
\qquad{} =\lambda_2(\bar z)h(\bv,\bar z)h(\bar z,\bar z) \sum \frac{f(\bet_{\so},\bet_{\st})g(\bxi_{\st},\bxi_{\so})}
{g(\bet_{\so},z_2)h(\bet_{\so},\bar z)h(\bxi_{\so}, \bar z)} \BV_{a+1,b}(\bet_{\st};\bxi_{\st}).\label{A6-T12T22}
\end{gather}
Here the sum is taken over partitions $\bet\Rightarrow\{\bet_{\so},\bet_{\st}\}$, $\bxi\Rightarrow\{\bxi_{\so},\bxi_{\st}\}$. The cardinalities of the subsets are $\#\bet_{\so}=1$, $\#\bxi_{\so}=2$.

\subsubsection[Successive action of $T_{22}$ and $T_{12}$]{Successive action of $\boldsymbol{T_{22}}$ and $\boldsymbol{T_{12}}$}

Using \eqref{ComRel2a} and \eqref{A6-T12T22} we are able to calculate the action of $T_{22}(z_1)T_{12}(z_2)$ onto $\BV_{a,b}(\bu;\bv)$. It is clear that for this we should take the following combination: equation \eqref{A6-T12T22} multiplied with $f(z_1,z_2)$ and the same equation with $z_1\leftrightarrow z_2$ multiplied with $g(z_2,z_1)$. This straightforward calculation gives
\begin{gather}\label{A1-T22T12}
T_{22}(z_1)T_{12}(z_2)\BV_{a,b}(\bu;\bv)=\lambda_2(\bar z)h(\bv,\bar z)h(\bar z,\bar z) \sum
\frac{f(\bet_{\so},\bet_{\st})g(\bxi_{\st},\bxi_{\so})} {h(\bet_{\so},z_1)h(\bxi_{\so}, \bar z)} \BV_{a+1,b}(\bet_{\st};\bxi_{\st}).
\end{gather}
Here the sum is taken over partitions $\bet\Rightarrow\{\bet_{\so},\bet_{\st}\}$, $\bxi\Rightarrow\{\bxi_{\so},\bxi_{\st}\}$.
The cardinalities of the subsets are $\#\bet_{\so}=1$, $\#\bxi_{\so}=2$.

\subsection{Recursion formula}

Now everything is ready for the use of recursion \eqref{recT12a}. Due to \eqref{recT12-n} we can write it as follows
\begin{gather}\label{recT12-nn}
\BV_{a+1,b}(\{\bar u,z_2\};\bar v)=\frac{1}{\lambda_2(z_2) f(\bv,z_2)}\bigl(T_{12}(z_2) \BV_{a,b}(\bar u;\bar v) - \Psi\bigr),
\end{gather}
where
\begin{gather}\label{Psi}
\Psi=\lambda_2(z_2)h(\bv,z_2)\sum_{z_2\notin \bxi_{\so}} \frac{g(\bxi_{\st},\bxi_{\so})}{h(\bxi_{\so},z_2)}
 \BV_{a+1,b}(\bet;\bxi_{\st}).
\end{gather}
Here $\bet=\{z_2,\bu\}$ and $\bxi=\{z_2,\bv\}$, and we used $h(z_2,z_2)=1$. The sum is taken over partitions $\bxi\Rightarrow\{\bxi_{\so}, \bxi_{\st}\}$ with $\#\bxi_{\so}=1$. One more restriction $z_2\notin \bxi_{\so}$ is shown explicitly by the subscript of the sum.

Recall that we assume that the action of $T_{22}(z_1)$ on the vectors $\BV_{a,b}(\bar u;\bar v)$ is given by~\eqref{A1-T22} at some value of $a\ge 0$ and arbitrary $b$. All the vectors in the linear combination~\eqref{Psi} have the form $\BV_{a+1,b}(\{\bu,z_2\};\{\bv_j,z_2\})$, that is $\{\bu,z_2\}\cap\{\bv_j,z_2\}\ne\varnothing$. Hence, taking into account Remark~\ref{rmk:T22T13}, the action of $T_{22}(z_1)$ on these vectors is known and it is given by~\eqref{A1-T22}:
\begin{gather}\label{T-Psi}
T_{22}(z_1)\Psi=\lambda_2(\bar z)h(\bv,z_2)\sum_{z_2\notin \bxi_{\so}} \frac{g(\bxi_{\st},\bxi_{\so})}{h(\bxi_{\so},z_2)}
h(\bxi_{\st},z_1)\frac{f(\bet_{\so},\bet_{\st})g(\bxi_{\rm ii},\bxi_{\rm i})}{h(\bet_{\so},z_1)h(\bxi_{\rm i}, z_1)}
\BV_{a+1,b}(\bet_{\st};\bxi_{\rm ii}).
\end{gather}
In this formula $\bet=\{\bar z,\bu\}$ and $\bxi=\{\bar z,\bv\}$. At the f\/irst step we have partitions $\{z_2,\bv\}\Rightarrow \{\bxi_{\so},\bxi_{\st}\}$. Then we obtain additional partitions $\{\bar z,\bv\}\Rightarrow\{\bxi_{\rm i},\bxi_{\rm ii}\}$ and $\bet\Rightarrow\{\bet_{\so},\bet_{\st}\}$. Hereby $\#\bet_{\so}=\#\bxi_{\so}=\#\bxi_{\rm i}=1$. Thus, one can say that the set $\bxi=\{\bar z,\bv\}$ is divided into subsets $\{\bxi_{\so},\bxi_{\rm i}, \bxi_{\rm ii}\}$ with the restrictions $z_1\notin\bxi_{\so}$ and $z_2\notin\bxi_{\so}$. Substituting $\bxi_{\st}=\{\bxi_{\rm i},\bxi_{\rm ii}\}\setminus \{z_1\}$ into \eqref{T-Psi} we obtain
\begin{gather}\label{T-Psi1}
T_{22}(z_1)\Psi=\lambda_2(\bar z)h(\bv,z_2)\sum_{z_2\notin \bxi_{\so}} \frac{g(\bxi_{\rm ii},\bxi_{\so})g(\bxi_{\rm i},\bxi_{\so})
g(\bxi_{\rm ii},\bxi_{\rm i})f(\bet_{\so},\bet_{\st})h(\bxi_{\rm ii},z_1)}
{g(z_1,\bxi_{\so})h(\bxi_{\so},z_2)h(\bet_{\so},z_1)}
\BV_{a+1,b}(\bet_{\st};\bxi_{\rm ii}).\!\!\!
\end{gather}
Observe that the restriction $z_1\notin\bxi_{\so}$ holds automatically due to the factor $g(z_1,\bxi_{\so})^{-1}$. In order to get rid of the restriction $z_2\notin \bxi_{\so}$ we present $T_{22}(z_1)\Psi$ as a dif\/ference of two terms. The f\/irst term is just the sum over partitions in~\eqref{T-Psi1}, where no restrictions on the partitions of the set~$\bxi$ are imposed. In the second term we simply set $\bxi_{\so}=z_2$. Thus,
\begin{gather*}
T_{22}(z_1)\Psi=\Psi'-\Psi'',
\end{gather*}
where
\begin{gather}\label{T-Psi-p}
\Psi'=\lambda_2(\bar z)h(\bv,z_2)\sum \frac{g(\bxi_{\rm ii},\bxi_{\so})g(\bxi_{\rm i},\bxi_{\so})
g(\bxi_{\rm ii},\bxi_{\rm i})f(\bet_{\so},\bet_{\st})h(\bxi_{\rm ii},z_1)}
{g(z_1,\bxi_{\so})h(\bxi_{\so},z_2)h(\bet_{\so},z_1)} \BV_{a+1,b}(\bet_{\st};\bxi_{\rm ii}),
\end{gather}
and
\begin{gather}\label{T-Psi-pp}
\Psi''=\lambda_2(\bar z)h(\bv,z_2)\sum \frac{g(\bxi_{\rm ii},z_2)g(\bxi_{\rm i},z_2)
g(\bxi_{\rm ii},\bxi_{\rm i})f(\bet_{\so},\bet_{\st})h(\bxi_{\rm ii},z_1)}
{g(z_1,z_2)h(\bet_{\so},z_1)} \BV_{a+1,b}(\bet_{\st};\bxi_{\rm ii}).
\end{gather}
In \eqref{T-Psi-pp} we have $\{\bxi_{\rm i},\bxi_{\rm ii}\}=\{\bv,z_1\}$, therefore
\begin{gather}\label{T-Psi-pp2}
\Psi''=\lambda_2(\bar z)h(\bv,z_1)f(\bv,z_2)\sum \frac{g(\bxi_{\rm ii},\bxi_{\rm i})f(\bet_{\so},\bet_{\st})}
{h(\bxi_{\rm i},z_1)h(\bet_{\so},z_1)}
\BV_{a+1,b}(\bet_{\st};\bxi_{\rm ii}).
\end{gather}

In \eqref{T-Psi-p} we can take the sum over partitions into subsets $\bxi_{\rm i}$ and $\bxi_{\so}$, because it consists of two terms only
\begin{gather*}
\frac{g(\bxi_{\rm i},\bxi_{\so})}{g(z_1,\bxi_{\so})h(\bxi_{\so},z_2)}+
\frac{g(\bxi_{\so},\bxi_{\rm i})}{g(z_1,\bxi_{\rm i})h(\bxi_{\rm i},z_2)}=
\frac{h(z_1,z_2)}{h(\bxi_0,z_2)},
\end{gather*}
where $\bxi_0=\{\bxi_{\rm i},\bxi_{\so}\}$. Thus,
\begin{gather}\label{T-Psi-p1}
\Psi'=\lambda_2(\bar z)h(\bv,z_2)h(z_1,z_2)\sum \frac{g(\bxi_{\rm ii},\bxi_{0})f(\bet_{\so},\bet_{\st})h(\bxi_{\rm ii},z_1)}
{h(\bxi_{0},z_2)h(\bet_{\so},z_1)} \BV_{a+1,b}(\bet_{\st};\bxi_{\rm ii}),
\end{gather}
and extracting the product $h(\bxi,z_1)$ we recast \eqref{T-Psi-p1} as follows
\begin{gather*}
\Psi'=\lambda_2(\bar z)h(\bv,\bar z)h(\bar z,\bar z)\sum \frac{g(\bxi_{\rm ii},\bxi_{0})f(\bet_{\so},\bet_{\st})}
{h(\bxi_{0},\bar z)h(\bet_{\so},z_1)} \BV_{a+1,b}(\bet_{\st};\bxi_{\rm ii}).
\end{gather*}
Here the sum is taken over partitions $\bet\Rightarrow\{\bet_{\so},\bet_{\st}\}$, $\bxi\Rightarrow\{\bxi_{0},\bxi_{\rm ii}\}$ with $\#\bet_{\so}=1$, $\#\bxi_{0}=2$. Comparing this expression with \eqref{A1-T22T12} we see that
\begin{gather*}
\Psi'=T_{22}(z_1)T_{12}(z_2)\BV_{a,b}(\bu;\bv).
\end{gather*}
Thus, we f\/ind from the recursion \eqref{recT12-nn}
\begin{gather*}
T_{22}(z_1)\BV_{a+1,b}(\{\bar u,z_2\};\bar v)=\frac{T_{22}(z_1)T_{12}(z_2) \BV_{a,b}(\bar u;\bar v) - \Psi'+\Psi''}
{\lambda_2(z_2) f(\bv,z_2)}=\frac{\Psi''}{\lambda_2(z_2) f(\bv,z_2)}.
\end{gather*}
Substituting \eqref{T-Psi-pp2} in this expression, we arrive at
\begin{gather*}
T_{22}(z_1)\BV_{a+1,b}(\{\bar u,z_2\};\bar v)=\lambda_2(z_1)h(\bv,z_1)\sum \frac{g(\bxi_{\st},\bxi_{\so})f(\bet_{\so},\bet_{\st})}
{h(\bxi_{\so},z_1)h(\bet_{\so},z_1)} \BV_{a+1,b}(\bet_{\st};\bxi_{\st}),
\end{gather*}
where we have relabeled $\bxi_{\rm i}\to \bxi_{\so}$ and $\bxi_{\rm ii}\to \bxi_{\st}$. Thus, the induction step is completed.

\subsection[Induction over $n$]{Induction over $\boldsymbol{n}$}

Actually, the induction over $n$ for the action of $T_{22}(\bar z)$ is a combination of the corresponding proofs for the actions of $T_{12}(\bar z)$ and $T_{23}(\bar z)$. Assume that \eqref{A-T22} is valid for some $n-1$. Then
\begin{gather*}
T_{22}(\bar z)\BV_{a,b}(\bu;\bv)=(-1)^{n-1}\lambda_2(\bar z_n)h(\bxi,\bar z_n)\\
\hphantom{T_{22}(\bar z)\BV_{a,b}(\bu;\bv)=}{}
\times \sum \frac{f(\bet_{\so},\bet_{\st})g(\bxi_{\st},\bxi_{\so})}{h(\bxi_{\so},\bar z_n)}K_{n-1}(\bar z_n|\bet_{\so}+c)
T_{22}(z_n)\BV_{a,b}(\bet_{\st};\bxi_{\st}).
\end{gather*}
Here the sum is taken over partitions $\{\bar z_n,\bv\}=\bxi\Rightarrow\{\bxi_{\so},\bxi_{\st}\}$ and $\{\bar z_n,\bu\}=\bet\Rightarrow\{\bet_{\so},\bet_{\st}\}$ with $\#\bxi_{\so}=\#\bet_{\so}=n-1$. Acting with $T_{22}(z_n)$ onto $\BV_{a,b}(\bet_{\st};\bxi_{\st})$ we obtain
\begin{gather*}
T_{22}(\bar z)\BV_{a,b}(\bu;\bv)=(-1)^{n}\lambda_2(\bar z)\frac{h(\bxi,\bar z_n)}{h(z_n,\bar z_n)}\sum
\frac{f(\bet_{\so},\bet_{\st})g(\bxi_{\st},\bxi_{\so})}{h(\bxi_{\so},\bar z_n)}K_{n-1}(\bar z_n|\bet_{\so}+c)\\
\hphantom{T_{22}(\bar z)\BV_{a,b}(\bu;\bv)=}{} \times h(\bxi_{\st},z_n)
\frac{f(\bet_{\rm i},\bet_{\rm ii})g(\bxi_{\rm ii},\bxi_{\rm i})}{h(\bxi_{\rm i},z_n)}K_{1}(z_n|\bet_{\rm i}+c)
\BV_{a,b}(\bet_{\rm ii};\bxi_{\rm ii}).
\end{gather*}
Here already $\bxi=\{\bar z,\bv\}$ and $\bet=\{\bar z,\bu\}$, and we have additional partitions $\{\bxi_{\st},z_n\}\Rightarrow\{\bxi_{\rm i},\bxi_{\rm ii}\}$ and $\{\bet_{\st},z_n\}\Rightarrow\{\bet_{\rm i},\bet_{\rm ii}\}$ with $\#\bxi_{\rm i}=\#\bet_{\rm i}=1$. Thus, we can say that we have the sum over partitions $\bxi\Rightarrow\{\bxi_{\so},\bxi_{\rm i},\bxi_{\rm ii}\}$ and $\bet\Rightarrow\{\bet_{\so},\bet_{\rm i},\bet_{\rm ii}\}$ with restrictions $z_n\notin \bet_{\so}$ and $z_n\notin \bxi_{\so}$.

Substituting $\bxi_{\st}=\{\bxi_{\rm i},\bxi_{\rm ii}\}\setminus \{z_n\}$, $\bet_{\st}=\{\bet_{\rm i},\bet_{\rm ii}\}\setminus \{z_n\}$ and denoting $\bxi_{0}=\{\bxi_{\rm i},\bxi_{\so}\}$, $\bet_{0}=\{\bet_{\rm i},\bet_{\so}\}$ we obtain
\begin{gather}
T_{22}(\bar z)\BV_{a,b}(\bu;\bv)=(-1)^{n}\lambda_2(\bar z)\frac{h(\bxi,\bar z)}{h(z_n,\bar z_n)}\sum
\frac{f(\bet_{\so},\bet_{\rm i})f(\bet_{0},\bet_{\rm ii}) g(\bxi_{\rm i},\bxi_{\so})g(\bxi_{\rm ii},\bxi_{0})}
{f(\bet_{\so},z_n) g(z_n,\bxi_{\so}) h(\bxi_{\so},\bar z_n)h(\bxi_{0},z_n)}\nonumber\\
\hphantom{T_{22}(\bar z)\BV_{a,b}(\bu;\bv)=}{} \times
K_{n-1}(\bar z_n|\bet_{\so}+c)K_{1}(z_n|\bet_{\rm i}+c) \BV_{a,b}(\bet_{\rm ii};\bxi_{\rm ii}).\label{A-T22n-3}
\end{gather}
Observe that the restrictions $z_n\notin \bet_{\so}$ and $z_n\notin \bxi_{\so}$ hold automatically due to the factors $f(\bet_{\so},z_n)$ and $g(z_n,\bxi_{\so})$ in the denominator of \eqref{A-T22n-3}. Using $K_{1}(z_n|\bet_{\rm i}+c)=-K_{1}(\bet_{\rm i}|z_n)/f(\bet_{\rm i},z_n)$ we recast~\eqref{A-T22n-3} in the form
\begin{gather*}
T_{22}(\bar z)\BV_{a,b}(\bu;\bv)=(-1)^{n-1}\lambda_2(\bar z)\frac{h(\bxi,\bar z)}{h(z_n,\bar z_n)}\sum
\frac{f(\bet_{0},\bet_{\rm ii}) g(\bxi_{\rm ii},\bxi_{0})}{f(\bet_{0},z_n) g(z_n,\bxi_{0}) h(\bxi_{0},\bar z)}\\
\qquad{}\times \bigl\{g(\bxi_{\rm i},\bxi_{\so})g(z_n,\bxi_{\rm i})h(\bxi_{\rm i},\bar z_n)\bigr\}
\bigl\{K_{n-1}(\bar z_n|\bet_{\so}+c)K_{1}(\bet_{\rm i}|z_n)f(\bet_{\so},\bet_{\rm i})\bigr\} \BV_{a,b}(\bet_{\rm ii};\bxi_{\rm ii}).
\end{gather*}
The sums over partitions $\bxi_0\Rightarrow\{\bxi_{\rm i},\bxi_{\so}\}$ and $\bet_0\Rightarrow\{\bet_{\rm i},\bet_{\so}\}$ (see the terms in braces) were already computed (see~\eqref{CI} and~\eqref{ML}). Thus, we arrive at
\begin{gather*}
T_{22}(\bar z)\BV_{a,b}(\bu;\bv)=(-1)^{n}\lambda_2(\bar z)h(\bxi,\bar z)\sum \frac{f(\bet_{0},\bet_{\rm ii}) g(\bxi_{\rm ii},\bxi_{0})}
{ h(\bxi_{0},\bar z)} K_{n}(\bar z|\bet_{0}+c) \BV_{a,b}(\bet_{\rm ii};\bxi_{\rm ii}),
\end{gather*}
which ends the proof.

\section[Induction over $n$ for the actions of $T_{ij}(\bar z)$ with $i>j$]{Induction over $\boldsymbol{n}$ for the actions of $\boldsymbol{T_{ij}(\bar z)}$ with $\boldsymbol{i>j}$}\label{Induction}

The action formulas for all other elements of the monodromy matrix can be proved exactly in the same manner. However, it is clear that the technical dif\/f\/iculty of the proofs increases when moving from the right top corner of the monodromy matrix to the left bottom corner. It is due to the form of the recursion formulas and the commutation relations~\eqref{commTT}. For example, we have seen that for the derivation of the action of $T_{22}(\bar z)$ one should know the actions of~$T_{12}(\bar z)$ and~$T_{23}(\bar z)$ onto Bethe vectors. The latest are relatively simple. However, one can easily convince oneself that to get the action of $T_{ij}(\bar z)$ with $i>j$, it is necessary to know the actions of the diagonal elements $T_{ii}(\bar z)$, which are more involved. Therefore, we omit the detailed proofs of the multiple actions of the operators $T_{11}(\bar z)$ \eqref{A-T11}, $T_{33}(\bar z)$ \eqref{A-T33}, and the operators $T_{ij}(\bar z)$ from the lower-triangular part of the monodromy matrix \eqref{A-T21}--\eqref{A-T31}. However, as an illustration of the method, we prove the multiple action of the operator $T_{21}(\bar z)$ assuming that the action of a~single operator~$T_{21}(z)$ is known.

As previously, the proof goes by induction over $n=\#\bar z$. We assume that the action \eqref{A-T21} holds for some $n-1$. Then acting successively with $T_{21}(\bar z_n)$ and $T_{21}(z_n)$ on $\BV_{a,b}(\bu;\bv)$ we obtain
\begin{gather}
T_{21}(\bar z)\BV_{a,b}(\bu;\bv)=\lambda_2(\bar z)\frac{h(\bxi,\bar z_n)}{h(z_n,\bar z_n)}\sum r_1(\bet_{\so})
\frac{f(\bet_{\st},\bet_{\so})f(\bet_{\st},\bet_{\sth})f(\bet_{\sth},\bet_{\so})g(\bxi_{\st},\bxi_{\so})}
{h(\bxi_{\so},\bar z_n)f(\bxi_{\st},\bet_{\so})}\nonumber\\
\qquad{} \times K_{n-1}(\bar z_n|\bet_{\st}+c)K_{n-1}(\bet_{\so}|\bxi_{\so}+c)
 h(\bxi_{\st},z_n) r_1(\bet_{\rm i})
\frac{f(\bet_{\rm ii},\bet_{\rm i})f(\bet_{\rm ii},\bet_{\rm iii})f(\bet_{\rm iii},\bet_{\rm i})g(\bxi_{\rm ii},\bxi_{\rm i})}
{h(\bxi_{\rm i},z_n)f(\bxi_{\rm ii},\bet_{\rm i})}\nonumber\\
\qquad{} \times K_1(z_n|\bet_{\rm ii}+c)K_1(\bet_{\rm i}|\bxi_{\rm i}+c)
\BV_{a-n,b}(\bet_{\rm iii};\bxi_{\rm ii}).\label{A1-T21T21}
\end{gather}
Here $\#\bet_{\rm i}=\#\bet_{\rm ii}=\#\bxi_{\rm i}=1$ and $\#\bet_{\so}=\#\bet_{\st}=\#\bxi_{\so}=n-1$. Originally we have partitions $\{\bar z_n,\bu\}=\bet\Rightarrow\{\bet_{\so},\bet_{\st},\bet_{\sth}\}$ and $\{\bar z_n,\bv\}=\bxi\Rightarrow\{\bxi_{\so},\bxi_{\st}\}$. Then we have additional partitions $\{z_n,\bet_{\st}\}\Rightarrow\{\bet_{\rm i},\bet_{\rm ii},\bet_{\rm iii}\}$ and $\{z_n,\bxi_{\st}\}\Rightarrow\{\bxi_{\rm i},\bxi_{\rm ii}\}$. Thus, in equation \eqref{A1-T21T21} we have $\{\bar z,\bu\}=\bet$ and $\{\bar z,\bv\}=\bxi$. Setting there $\bet_{\sth}= \{\bet_{\rm i},\bet_{\rm ii},\bet_{\rm iii}\}\setminus \{z_n\}$ and $\bxi_{\st}=\{\bxi_{\rm i},\bxi_{\rm ii}\}\setminus \{z_n\}$ we arrive at
\begin{gather}
T_{21}(\bar z)\BV_{a,b}(\bu;\bv)=\lambda_2(\bar z)\frac{h(\bxi,\bar z)}{h(z_n,\bar z_n)}\sum r_1(\bet_{\so})r_1(\bet_{\rm i})
\frac{f(\bet_{\rm ii},\bet_{\rm i})f(\bet_{\rm ii},\bet_{\rm iii})f(\bet_{\rm iii},\bet_{\rm i})
g(\bxi_{\rm ii},\bxi_{\rm i})}
{h(\bxi_{\rm i},z_n)f(\bxi_{\rm ii},\bet_{\rm i})}\nonumber\\
\qquad{}
\times\frac{f(\bet_{\st},\bet_{\so})f(\bet_{\st},\bet_{\rm i})f(\bet_{\st},\bet_{\rm ii})f(\bet_{\st},\bet_{\rm iii})
f(\bet_{\rm i},\bet_{\so})f(\bet_{\rm ii},\bet_{\so})f(\bet_{\rm iii},\bet_{\so})g(\bxi_{\rm i},\bxi_{\so})g(\bxi_{\rm ii},\bxi_{\so})}
{h(\bxi_{\so},z_n)h(\bxi_{\so},\bar z_n)
f(\bxi_{\rm i},\bet_{\so})f(\bxi_{\rm ii},\bet_{\so})f(\bet_{\st},z_n)g(z_n,\bxi_{\so})}\nonumber\\
\qquad{} \times K_{n-1}(\bar z_n|\bet_{\st}+c)K_1(z_n|\bet_{\rm ii}+c)K_{n-1}(\bet_{\so}|\bxi_{\so}+c)K_1(\bet_{\rm i}|\bxi_{\rm i}+c)
\BV_{a-n,b}(\bet_{\rm iii};\bxi_{\rm ii}).\label{A2-T21T21}
\end{gather}
Now we set $\{\bet_{\so},\bet_{\rm i}\}=\bet_{0}$, $\{\bet_{\st},\bet_{\rm ii}\}=\bet_{0'}$, and $\{\bxi_{\so},\bxi_{\rm i}\} =\bxi_{0}$. We also transform $K_1(z_n|\bet_{\rm ii}+c)=-K_1(\bet_{\rm ii}|z_n)/f(\bet_{\rm ii},z_n)$ and $K_1(\bet_{\rm i}|\bxi_{\rm i}+c)=-K_1(\bxi_{\rm i}|\bet_{\rm i})/f(\bxi_{\rm i},\bet_{\rm i})$. Then \eqref{A2-T21T21} takes the form
\begin{gather}
T_{21}(\bar z)\BV_{a,b}(\bu;\bv)=\lambda_2(\bar z)\frac{h(\bxi,\bar z)}{h(z_n,\bar z_n)}\sum
\frac{r_1(\bet_{0})f(\bet_{0'},\bet_{0})f(\bet_{0'},\bet_{\rm iii})f(\bet_{\rm iii},\bet_{0})
g(\bxi_{\rm ii},\bxi_{0})g(\bxi_{\rm i},\bxi_{\so})}
{h(\bxi_{\so},\bar z) f(\bxi_{\rm i},\bet_{0})f(\bet_{0'},z_n)
f(\bxi_{\rm ii},\bet_{0}) g(z_n,\bxi_{\so})h(\bxi_{\rm i},z_n)}\nonumber\\
\qquad{} \times \big\{ K_{n-1}(\bar z_n|\bet_{\st}+c)K_1(\bet_{\rm ii}|z_n)f(\bet_{\st},\bet_{\rm ii})\big\}
\big\{ K_{n-1}(\bet_{\so}|\bxi_{\so}+c)K_1(\bxi_{\rm i}|\bet_{\rm i})f(\bet_{\rm i},\bet_{\so})\big\}\nonumber\\
\qquad{}\times
\BV_{a-n,b}(\bet_{\rm iii};\bxi_{\rm ii}).\label{A3-T21T21}
\end{gather}
The sums over partitions in braces can be computed via \eqref{Sym-Part-old1}:
\begin{gather*}
\sum_{\bet_{0'}\Rightarrow\{\bet_{\st},\bet_{\rm ii}\}}K_{n-1}(\bar z_n|\bet_{\st}+c)K_1(\bet_{\rm ii}|z_n)f(\bet_{\st},\bet_{\rm ii})=
-f(\bet_{0'},z_n)K_n(\bar z |\bet_{0'}+c),
\end{gather*}
and
\begin{gather*}
\sum_{\bet_{0}\Rightarrow\{\bet_{\so},\bet_{\rm i}\}} K_{n-1}(\bet_{\so}|\bxi_{\so}+c)K_1(\bxi_{\rm i}|\bet_{\rm i})f(\bet_{\rm i},\bet_{\so})
=(-1)^{n-1}\frac{K_n(\bxi_{0}|\bet_{0})}{f(\bxi_{\so},\bet_{0})}.
\end{gather*}
Substituting this into \eqref{A3-T21T21} we arrive at
\begin{gather*}
T_{21}(\bar z)\BV_{a,b}(\bu;\bv)=\lambda_2(\bar z)\frac{h(\bxi,\bar z)}{h(z_n,\bar z_n)}\sum
\frac{r_1(\bet_{0})f(\bet_{0'},\bet_{0})f(\bet_{0'},\bet_{\rm iii})f(\bet_{\rm iii},\bet_{0})
g(\bxi_{\rm ii},\bxi_{0})} {h(\bxi_{0},\bar z) f(\bxi_{\rm ii},\bet_{0}) g(z_n,\bxi_{0})}\\
\hphantom{T_{21}(\bar z)\BV_{a,b}(\bu;\bv)=}{}
\times K_n(\bar z |\bet_{0'}+c) K_n(\bet_{0}|\bxi_{0}+c)
\bigl\{ g(\bxi_{\rm i},\bxi_{\so})h(\bxi_{\rm i},\bar z_n)g(z_n,\bxi_{\rm i})\bigr\}
\BV_{a-n,b}(\bet_{\rm iii};\bxi_{\rm ii}),
\end{gather*}
where we again replaced
$K_n(\bxi_{0}|\bet_{0})$ by $K_n(\bet_{0}|\bxi_{0}+c)$ via \eqref{K-K}. The sum over partitions in braces can be calculated via Lemma~\ref{main-ident-C}
\begin{gather*}
\sum_{\bxi_{0}\Rightarrow\{\bxi_{\so},\bxi_{\rm i}\}} g(\bxi_{\rm i},\bxi_{\so})h(\bxi_{\rm i},\bar z_n)g(z_n,\bxi_{\rm i})=
h(z_n,\bar z_n)g(z_n,\bxi_{0}).
\end{gather*}
Thus, we f\/inally obtain
\begin{gather*}
T_{21}(\bar z)\BV_{a,b}(\bu;\bv)=\lambda_2(\bar z)h(\bxi,\bar z)\sum
\frac{r_1(\bet_{0})f(\bet_{0'},\bet_{0})f(\bet_{0'},\bet_{\rm iii})f(\bet_{\rm iii},\bet_{0})
g(\bxi_{\rm ii},\bxi_{0})} {h(\bxi_{0},\bar z) f(\bxi_{\rm ii},\bet_{0})}\\
\hphantom{T_{21}(\bar z)\BV_{a,b}(\bu;\bv)=}{}
\times K_n(\bar z |\bet_{0'}+c) K_n(\bet_{0}|\bxi_{0}+c) \BV_{a-n,b}(\bet_{\rm iii};\bxi_{\rm ii}),
\end{gather*}
which coincides with the original formula up to the labeling of the subsets.

\section{Conclusion}

In this paper we obtained compact expressions for the multiple actions of the operators $T_{ij}$ onto the Bethe vectors in the models with~$\mathfrak{gl}(2|1)$ symmetry. Formally, our method can be extended to higher rank algebras as well, although the technical complexity of the calculations increases signif\/icantly with the rank and can be rapidly untractable.

Comparing the actions of $T_{ij}$ onto $\BV_{a,b}(\bu;\bv)$ with the analogous actions in the $\mathfrak{gl}(3)$-based models \cite{BelPRS13a} one can observe that they are quite similar. The main dif\/ference is that some of the Izergin determinants in the~$\mathfrak{gl}(3)$ actions are replaced with the Cauchy determinants in their~$\mathfrak{gl}(2|1)$ analogs. This is the consequence of~$\mathbb{Z}_2$ grading that `trivializes' some of the relations and helps to go further in the calculation. It is well possible that there exist some general formulas, which valid for both cases. It would be important to f\/ind such formulas, especially for the models with graded algebras of higher rank.

The obtained formulas allow one to tackle the problem of scalar products of the Bethe vectors. For this we can use explicit expressions for the dual Bethe vectors $\mathbb{C}_{a,b}(\bu,\bv)$ obtained in \cite{PakRS16a}. Actually, they can be obtained by a transposition\footnote{More precisely, one should use an antimorphism of the algebra $\mathfrak{gl}(2|1)$ relating $T_{ij}$ and $T_{ji}$ (see~\cite{PakRS16a} for details).}
of \eqref{Phi-expl1}, \eqref{Phi-expl2}, for instance,
\begin{gather*}
\mathbb{C}_{a,b}(\bu,\bv)=(-1)^{b(b-1)/2}\sum g(\bv_{\so},\bu_{\so})
\frac{f(\bu_{\so},\bu_{\st}) g(\bv_{\st},\bv_{\so})h(\bu_{\so},\bu_{\so})}{\lambda_2(\bv_{\st})\lambda_2(\bu) f(\bv,\bu)}
\Omega^\dagger \bT_{32}(\bv_{\st}) T_{21}(\bu_{\st})\bT_{31}(\bu_{\so}).
\end{gather*}
Having this representation we reduce the evaluation of the scalar product to the calculation of successive multiple actions of the operators $T_{31}$, $T_{21}$, and $T_{32}$ onto Bethe vectors. This will be the subject of our further publication.

\appendix

\section{Identities for rational functions \label{A-ID}}

\begin{Lemma}\label{main-ident-C}
Let $\bw$, $\bu$ and $\bv$ be sets of complex variables with $\#\bu=m_1$,
$\#\bv=m_2$, and $\#\bw=m_1+m_2$, where $m_1$ and $m_2$ are fixed arbitrary integers. Then
\begin{gather}\label{Sym-Part-new1}
 \sum g(\bw_{\so},\bu)g(\bw_{\st},\bv)g(\bw_{\st},\bw_{\so}) = \frac{g(\bw,\bu)g(\bw,\bv)}{g(\bu,\bv)}.
\end{gather}
The sum is taken with respect to all partitions of the set $\bw$ into subsets $\bw_{\so}$ and $\bw_{\st}$ with $\#\bw_{\so}=m_1$ and $\#\bw_{\st}=m_2$.
\end{Lemma}

The proof of this lemma is given in \cite{Sla16}. Let us show how Lemma~\ref{main-ident-C} works. In equation~\eqref{CI} we have a sum
\begin{gather*}
I=\sum_{\bxi_{0}\Rightarrow \{\bxi_{\so},\bxi_{\rm i}\}}
 g(\bxi_{\rm i},\bxi_{\so})g(z_n,\bxi_{\rm i})h(\bxi_{\rm i},\bar z_n),
\end{gather*}
where $\#\bxi_{\rm i}=1$ and $\#\bxi_{\so}=n-1$. First, we reduce this sum to the form \eqref{Sym-Part-new1} using $h(u,v)=1/g(u,v-c)$. We have
\begin{gather*}
I=-h(\bxi_{0},\bar z_n)\sum_{\bxi_{0}\Rightarrow \{\bxi_{\so},\bxi_{\rm i}\}}
 g(\bxi_{\rm i},\bxi_{\so})\frac{g(\bxi_{\rm i},z_n)}{h(\bxi_{\so},\bar z_n)}=
 -h(\bxi_{0},\bar z_n)\sum_{\bxi_{0}\Rightarrow \{\bxi_{\so},\bxi_{\rm i}\}}
 g(\bxi_{\rm i},\bxi_{\so})g(\bxi_{\rm i},z_n)g(\bxi_{\so},\bar z_n-c).
\end{gather*}
Now we can directly apply \eqref{Sym-Part-new1}, and we arrive at
\begin{gather*}
I=-h(\bxi_{0},\bar z_n)\frac{g(\bxi_{0},z_n)g(\bxi_{0},\bar z_n-c)}{g(\bar z_n-c,z_n)}=g(z_n,\bxi_{0})h(z_n,\bar z_n).
\end{gather*}

\begin{Lemma}\label{main-ident}
Let $\bw$, $\bu$ and $\bv$ be sets of complex variables with $\#\bu=m_1$, $\#\bv=m_2$, and $\#\bw=m_1+m_2$. Then
\begin{gather}\label{Sym-Part-old1}
 \sum K_{m_1}(\bw_{\so}|\bu)K_{m_2}(\bv|\bw_{\st})f(\bw_{\st},\bw_{\so}) = (-1)^{m_1}f(\bw,\bu) K_{m_1+m_2}(\{\bu-c,\bv\}|\bw).
\end{gather}
The sum is taken with respect to all partitions of the set $\bw$ into subsets $\bw_{\so}$ and $\bw_{\st}$ with $\#\bw_{\so}=m_1$ and $\#\bw_{\st}=m_2$.
\end{Lemma}

The proof of this lemma is given in~\cite{BelPRS12a}.

\subsection*{Acknowledgements}
The work of A.L.~has been funded by the Russian Academic Excellence Project 5-100 and by joint NASU-CNRS project F14-2016. The work of S.P.~was supported in part by the RFBR grant 16-01-00562-a. N.A.S.~was supported by the grants RFBR-15-31-20484-mol-a-ved and RFBR-14-01-00860-a.

\pdfbookmark[1]{References}{ref}
\LastPageEnding

\end{document}